\documentclass[twocolumn,amsmath,amssymb,aps,pre,superscriptaddress]{revtex4}
\usepackage{color}
\usepackage{cancel}
\usepackage{comment}
\usepackage{graphicx}
\usepackage{dcolumn}
\usepackage{enumerate}
\usepackage{bm}
\usepackage{hyperref}

\usepackage[colorinlistoftodos]{todonotes}

\newcommand{\changed}[1]{ #1}

\def\widthprop{27.5mm}
\def\widthvertex{22.5mm}

\newcommand{\SA}{\mathcal{S}}

\newcommand{\eps}{\varepsilon}

\newcommand{\mx}{{\bm x}}
\newcommand{\mk}{{\bm k}}

\newcommand{\mpp}{{\bm p}}
\newcommand{\boldnabla}{{\bm \nabla}}

\newcommand{\tpsi}{ \tilde{\psi}}

\newcommand{\tphi}{ \tilde{\phi}}

\usepackage{longtable}

\allowdisplaybreaks 

\begin{document}


\title{Field-theoretic Analysis of Dynamic Isotropic Percolation: Three-loop Approximation}


\author{Michal Hnati\v{c}}%
\email{hnatic@saske.sk}
\affiliation{Bogolyubov Laboratory of Theoretical Physics, Joint Institute for Nuclear Research, 141980 Dubna, Russian Federation}
\affiliation{Institute of Experimental Physics, Slovak Academy of Sciences, Watsonova 47, 040 14 Ko\v{s}ice, Slovakia}
\affiliation{Faculty of Science, \v{S}af\'{a}rik University, Moyzesova 16, 040 01 Ko\v{s}ice, Slovakia}

\author{Matej Kecer}
\email{matej.kecer@student.upjs.sk}
\affiliation{Faculty of Science, \v{S}af\'{a}rik University, Moyzesova 16, 040 01 Ko\v{s}ice, Slovakia}

%
%

\author{Mikhail V. Kompaniets}
\email{m.kompaniets@spbu.ru}
\affiliation{Bogolyubov Laboratory of Theoretical Physics, Joint Institute for Nuclear Research, 141980 Dubna, Russian Federation}
\affiliation{Sankt Petersburg State University, St. Petersburg 199034, Russian Federation}

\author{Tom\'{a}\v{s} Lu\v{c}ivjansk\'{y}}
\email{tomas.lucivjansky@upjs.sk}
\affiliation{Faculty of Science, \v{S}af\'{a}rik University, Moyzesova 16, 040 01 Ko\v{s}ice, Slovakia}

\author{Luk\'{a}\v{s} Mi\v{z}i\v{s}in}
\email{mizisin@theor.jinr.ru}
\affiliation{Bogolyubov Laboratory of Theoretical Physics, Joint Institute for Nuclear Research, 141980 Dubna, Russian Federation}

\author{Yurii \,G.\,Molotkov}
\email{molotkov@theor.jinr.ru}
\affiliation{Bogolyubov Laboratory of Theoretical Physics, Joint Institute for Nuclear Research, 141980 Dubna, Russian Federation}
%
%

\date{\today}

\begin{abstract}
The general epidemic process is a paradigmatic model in non-equilibrium statistical physics displaying a continuous phase transition between active and absorbing states.
 The dynamic isotropic percolation universality class captures its universal properties, which
 we aim to quantitatively study by means of
   the field-theoretic formulation of the model augmented with a perturbative renormalization group analysis. 
  The main purpose of this work consists 
  in determining the critical dynamic exponent $z$ to the three-loop approximation. This allows us to finalize the quantitative description of the dynamic isotropic percolation class to this order of perturbation theory.
 The calculations are performed within the dimensional regularization with
  the minimal subtraction scheme and actual perturbative expansions are carried out in a formally small parameter $\eps$, where 
  $\eps = 6-d$
  is a
   deviation from the upper critical dimension $d_c=6$.    
 \end{abstract}

\maketitle


\section{\label{sec:level1} Introduction}

Non-equilibrium processes are prevalent in nature and the majority of observed phenomena are in 
 some form of non-equilibrium state~\cite{marro_dickman1999,krapivsky_book2010}. 
 Variety of related phenomena encompass turbulent flows~\cite{davidson}, pattern formations~\cite{hohenberg1993}, Earth's atmosphere~\cite{marston2012}, and living organisms~\cite{wang2019}.
 A plethora of other examples can be found not only in the realm of physics and biology but also in chemistry, economy, sociology, climatology, and other research areas. The pervasiveness of the non-equilibrium systems raises the importance of their understanding, which
  is of utmost importance both for theoretical and practical applications. 

From a whole set of possible non-equilibrium problems, an important collection is formed by growth models, which
find a lot of applications in population dynamics, the creation of fractal structures, etc.
 Theoretically, such models can be described as
stochastic systems in which the microscopic degrees of freedom evolve according to prescribed probabilistic rules. Under specific circumstances it often happens that the collective
 behavior of many microscopic entities 
 allows us to employ continuum approximation in which certain
 non-universal properties, such as the type of a lattice structure, are absent. On the other hand, non-trivial
 universal behavior is manifested in various quantities
 such as critical exponents \cite{Tauber2014}.
 We anticipate such behavior whenever a dynamical system undergoes a continuous phase transition akin to equilibrium second-order phase transitions.
In this paper, our aim is to study the universal properties of the general epidemic process (GEP).
  The model is also known in the literature as an epidemic with (full) recovery.
In the lattice formulation of this model, 
each site of a regular lattice can be in one of three possible states. These correspond to susceptible, infected, and immune individuals. Customarily, it is assumed that initially the outbreak is spatially localized, i.e., all sites are in a susceptible state and a single
infected seed is placed at the 
origin \cite{Henkel2008}. 
 Infected individuals then locally spread the infection 
  to neighboring sites. In this way, the model stochastically
 evolves in time. The crucial property
 of the GEP lies in the effect of immunization, i.e.,
  an infected site becomes at a certain rate immune and
  does not spread the disease further. 
  The immunization process is assumed to be perfect, i.e., a recovered site cannot be infected again. 
  The value of the infection rate crucially affects
  the macroscopic (large spatial and temporal scales) behavior of the GEP. If the infection
rate is low, we expect the 
disease to be spread only
in a local neighborhood of an initial seed that in biological terms corresponds to the endemic spreading. 
After a finite amount of time 
the activity dies out,
leaving a certain limited cluster of immune sites behind. This is known in the literature \cite{Henkel2008,Tauber2014} as the absorbing (inactive) phase of the model. In the opposite limit of large infection rates, there is a
finite probability that number of infected sites escalates
system-wide in the form of an expanding front of active sites. 
This corresponds to the so-called active 
%
%
state. 
Both
(active and absorbing) phases 
are separated by a non-equilibrium phase transition where, 
 at criticality, underlying microscopic degrees of freedom behave collectively over many spatiotemporal scales. 
 This gives rise to an intriguing self-similar scaling
 behavior, whose particular hallmark is a large correlation length with respect to both time and spatial directions.
 It can then be argued that this directly permits an approximation of the system by means of the continuous fields \cite{cardy1996scaling,Zinn2002,Zinn2007}.
  Averaging over small volumes, whose diameter is much smaller than the correlation length, leads to an effective description in terms of field variables.
  It is well-known~\cite{Zinn2007,Kardar2007} that such
  a mesoscopic approach
  facilitates the use of powerful methods of statistical field theory.
Further, many concepts and methods from equilibrium critical models can be taken over and used for theoretical analysis. One of the most important is 
 the concept of universality class. According to it, systems can be categorized into different classes, whose members share the same
 macroscopic behavior. Completely different systems from a microscopic point of view
 could display the same critical behavior, which is determined to a great extent
  by some common gross properties such as space dimension, symmetry, nature of order parameter, etc. All systems in the same universality class are quantitatively
   described by the same set of critical exponents. Therefore, to analyze their critical behavior it is advantageous to choose the simplest possible
  member of a given class. The phase transition in GEP belongs to the universality class of dynamic isotropic percolation 
%
%
\changed{(dIP)}
%
%
  \cite{Grassberger1983, Janssen1985, Cardy1985, Janssen2005} and hence it is this model that needs to be examined.

 In this paper, we aim to quantitatively analyze 
 the universality class of 
%
%
\changed{dIP}
%
%
using the field-theoretic renormalization group up to the three-loop approximation. This paper essentially consists of three
  major steps.
First, we formulate a dynamic isotropic percolation process in the parlance of the functional integrals.
Though far from being the decisive approach to
a problem, this method offers great insight into the origin of universality and validation of scaling relations. 
Second, the ensuing action functional is amenable to perturbative calculations in which universal quantities can be systematically estimated \cite{Vasilev2004}. 
Accordingly, we calculate perturbatively relevant Green functions conveniently expressed in the form of Feynman diagrams~\cite{Zinn2007,ChaikinLubensky}.
Third, divergent Feynman diagrams are treated with
 a perturbative renormalization group (RG), which not only furnishes a conceptual formalism but also
   provides a powerful and versatile mechanism for computing universal quantities such as critical exponents, amplitudes, and even perturbative calculation of scaling functions~\cite{Vasilev2004}. 
   Once the model is properly renormalized, we
    gain information about possible large-scale behavior.
   An important parameter in the RG procedure is the upper 
  critical dimension $d_c$ (Here, $d_c=6$). Above $d_c$, the mean-field theory, which completely neglects spatial fluctuations of the order parameter, predicts correct values for the critical exponents~\cite{Amit,Vasilev2004,Zinn2007}. 
   On the other hand, below $d_c$, fluctuations dominate the behavior of the critical system and the mean-field approximation is not valid anymore. 
 At the critical dimension $d_c$, RG theory predicts mean-field results with logarithmic corrections~\cite{Vasilev2004,Janssen2003}.
    
To treat ultraviolet (UV) divergences in Feynman graphs
 we employ dimensional regularization combined with 
 $\eps$-expansion~\cite{Zinn2002,Vasilev2004}.
 Using the former method, we effectively regularize divergent Feynman diagrams, whereas the latter method is a convenient way for perturbative calculations of physical expressions~\cite{Zinn2002,Vasilev2004}.
 Formally, a small $\eps$-expansion parameter is given by a difference $d_c-d$ from the upper critical dimension (as we will see later on in Sec.~\ref{subsec:can_dim})
 $d_c=6$ for 
%
%
\changed{dIP}
%
%
 process). The main quantitative results of RG are asymptotic series for critical exponents, and these have to be properly 
  summed to get precise numerical estimations~\cite{BenderBook}. 
  Currently, to the best of our knowledge, analytical predictions for static critical exponents of the universality class of isotropic percolation are known up
  to the fifth-order of the perturbation theory
  \cite{borinsky2021}. 
  However, there still exists a missing piece even at the level of three loops, which is related to the dynamic exponent $z$ governing scaling
  behavior in a temporal direction. In contrast to static
   critical exponents, the determination of the exponent $z$ requires
  the use of the full dynamic functional \cite{Janssen2005}.
  This, however, poses non-trivial technical obstacles in actual
  calculations. Our aim in this paper is to extend the existing perturbative results and finalize a complete three-loop
  analysis of dynamic isotropic percolation.
 
 Let us stress that multi-loop calculations are usually
accompanied by technically demanding tasks related to
both symbolic and numeric calculations.
Typically, perturbative calculations
are analytically feasible only for a small number of loops. 
In fact, the majority of calculations for dynamic models (including non-equilibrium models)
are limited only to the two-loop approximations  \cite{Vasilev2004,Tauber2014}. 
The third-order perturbation theory 
thus by itself poses a non-trivial improvement of existing results. On the other hand, knowledge about high-order 
terms offers further information on the behavior of perturbative results through Pad\'e (or other related) resummation schemes.

It might be argued \cite{adzhemyan2003} that the
complexity of three-loop calculations in dynamic models is
 more than an order of magnitude 
higher than that
 of the two-loop approximation. 
 Not only is the number of Feynman diagrams much higher, but there are also numerical problems with the correct
 extraction of divergent parts of Feynman diagrams.
 In this paper, we combine both analytical and numerical techniques to calculate all necessary renormalization constants.
 Once the Feynman diagrams are calculated, we then apply the field-theoretic RG method in a semi-analytical fashion.

This paper is organized as follows. In Sec.~\ref{sec:DIP} the field-theoretic formulation of the isotropic percolation process is presented.
General aspects of the renormalization group analysis
are discussed in Sec~\ref{sec:RG}. Sec.~\ref{sec:com_rg_const} describes non-trivial aspects of RG calculations at the three-loop level and contains the main results of the paper in the form of critical exponents.
 Sec. \ref{sec:conclusion} is reserved for concluding remarks. 
 In supplementary material~\cite{supplement} we describe in detail
 various technical and numerical data about Feynman diagrams, their representation, algebraic structure, and divergent parts, which allows direct verification of our results.


{\section{Field-theoretic formulation of dynamic isotropic Percolation}
\label{sec:DIP}}
{\subsection{Response functional} \label{subsec:action}}
Let us briefly describe a mesoscopic formulation of
%
%
\changed{the dIP}
%
%
 process \cite{Janssen2005,Cardy1983,Janssen1985}, which is based on general phenomenological considerations - setting up stochastic equations of motion and later constructing De Dominicis-Janssen action functional \cite{Janssen1976,Dominicis1976}.
We summarize the main points of this approach as concisely formulated in the review paper \cite{Janssen2005}.
The guiding principles within the phenomenological approach can be effectively expressed in the epidemic formulation, and are as follows

%
%
\changed{
\begin{enumerate}[(1)]   
  
  \item The susceptible medium is locally infected, and the corresponding rate depends solely on the local density of active sites $\psi$ and accumulated density of the inactive debris $\varphi$ (defined later in Eq.~\eqref{eq:phi_def}), respectively.
   After a short time interval, the infected sites recover and form immobile immune debris. 

  \item The states with no active sites, i.e., 
  $\psi= 0$, and arbitrary configuration of $\varphi$ are considered absorbing. For these states, the disease is effectively extinct and the percolation process stops. In this sense, the GEP has infinitely many absorbing states \cite{Henkel2008}
and can be used as a dynamic prescription or as an algorithmic tool to grow
ordinary percolation clusters.
 
  \item The disease spreads out via a diffusion process facilitated by neighboring susceptible regions. The left debris
  does not spread any more.
  
  \item Fast microscopic degrees of freedom are accounted for by stochastic noise, which respects absorbing condition.
  
\end{enumerate}
}
%
%
Taking into account these principles the path to a phenomenological formulation of the 
%
%
\changed{dIP}
%
%
 process in terms of 
 the coarse-grained field $\psi = \psi(t,\mx)$ is straightforward \cite{Janssen2005,Janssen1985}, with $t$ is a time variable and $\mx = (x_1,\ldots,x_d)$ being $d$-dimensional Euclidean vector. 
 Let us note, that we retain this general $d$-dependence
 in all formulas since, in what follows, our technical tool
 consists of an application of dimensional regularization method, which is based on the underlying assumption that $d$ is a complex variable \cite{Vasilev2004,Zinn2007}.
Moreover, this provides a transparent way for independent verification of obtained results.
 
Universal features of the
%
%
\changed{dIP}
%
%
process can be modeled
by dynamic field $\psi(t,\mx)$ obeying the
coarse-grained Langevin equation 
\begin{equation}
  \partial_t \psi = D_0 (\boldnabla^2 - \tau_0) \psi  -  D_0 u_0 \psi \varphi  + \sqrt{\psi}\zeta ,  
  \label{eq:basic_eom}
\end{equation} 
where $\zeta = \zeta(t,\mx)$ is the local noise
term, $\partial_t = \partial / \partial t$ is
the time derivative, $\boldnabla^2 = \sum_{i=1}^d \partial^2/\partial x_i^2$ is the Laplace operator
 in $d$-spatial dimensions, $D_0$ 
is the diffusion constant, $u_0$ is the coupling constant and $\tau_0$ measures
 a deviation from the critical value for the injected probability
  (similar to a deviation from critical temperature 
  $\tau = T-T_c$   in famous $\varphi^4$-theory in critical statics \cite{Zinn2007}).
Let us note that recently there have been attempts
related to relaxing the assumption (5) mentioned above
\cite{deng2023,Nettuno2024}, where
 it was additionally assumed that the debris field was subject to further diffusion spreading.

Hereinafter, we denote unrenormalized quantities with the subscript ``0'', whereas renormalized quantities will be written without the subscript. 
The field corresponding to stationary inactive debris 
$\varphi$ is given by the time integral
\begin{equation}
  \varphi(t,\mx) = D_0\int\limits_{-\infty}^{t} dt'\, \psi(t', \mx),
  \label{eq:phi_def}
\end{equation}
which is based on an implicit assumption of a perfect
immunization.
The noise field $\zeta(t,\mx)$ is assumed to be a Gaussian random variable with zero mean and the correlation function taking the following form
\begin{equation}
  \left\langle \zeta(t, \mx) \zeta (t', \mx') \right\rangle =
  u_0 D_0 
  \delta(t-t') \delta^{(d)} ( \mx - \mx'),
  \label{eq:noise}
\end{equation}
where brackets $\langle\ldots\rangle$ stand for the averaging over
all permissible stochastic realizations, and $\delta^{(d)}(\mx)$ is $d$-dimensional realization of Dirac delta function.

Let us note that
the whole noise term in the Eq.~
\eqref{eq:basic_eom}
is known in the literature as multiplicative noise \cite{kampen2011} and obviously respects the absorbing condition of the model.

In order to apply field-theoretical methods it is advantageous to recast stochastic problem~\eqref{eq:basic_eom} in the functional formulations. This can be done in a well-established manner
and the ensuing De Dominicis-Janssen action functional \cite{Janssen1976,Dominicis1976, Tauber2014} captures universal properties of 
%
%
\changed{the dIP}
%
%
class. It takes the following form 
\begin{align}
  \SA & =  \int dx\, \biggl[ \tpsi\left( - \partial_t + D_0 \boldnabla^2 - D_0 \tau_0 \right) \psi 
  \nonumber\\
  & + \frac{u_0 D_0}{2}
   \left( \tpsi^2 \psi - 2 \tpsi \psi \varphi \right)
%
%
\changed{+k\tpsi}
%
%
   \biggl],
   \label{eq:action}
\end{align}
where for brevity we have written $x=(t,\mx)$, 
 $dx = dt d^d x$ is a shorthand for the integration measure, and 
$\tpsi = \tpsi(t,\mx)$ is the auxilliary
Martin-Siggia-Rose response field \cite{MSR1973,Vasilev2004},  
%
%
\changed{and $k$ stands for a source term (seed) which is usually chosen to correspond to an
initial sick individual located at $(t,\mx) = (0,\bm{0})$, i.e. $k$ is proportional
 to $\propto \delta(t)\delta(\mx)$ \cite{Janssen2005,Henkel2008}. From non-equilibrium perspective, such term is necessary for
an initialization of the spreading process, but apart from that it has no additional effect. 
Since this term does not affect
the renormalization procedure and hence
 also the properties of the model at the IR limit (long-time large-space) considered in present paper, this term will be suppressed.
For brevity, we also suppress the dependence on the variable $x$ in all entering fields $\psi,\tpsi$ and $\varphi$, respectively.}
%
%
 
 It is reassuring that it is also possible to arrive at the 
 same field-theoretic action \eqref{eq:action} by independent means.
  Starting with a specific reaction-diffusion system
   on a discrete lattice, one can employ
    Doi-Peliti field theory \cite{Doi1976a, Peliti1985} to derive an action in terms
    of bosonic-like fields \cite{Tauber2005,Tauber2014}. All involved
    theoretical steps are well-controlled, and,
   in the end, one obtains the action functional
   that possesses asymptotically equivalent properties to those of the action \eqref{eq:action}. 

By short inspection, we observe that the model described by the action \eqref{eq:action} at the
mean-field level (and simultaneously neglecting
noise variable) indeed exhibits 
 a phase transition between the active phase (with a nonzero value of $\psi$) and the absorbing phase (with the vanishing field $\psi$, i.e., $\psi = 0$).

The field-theoretic action \eqref{eq:action} exhibits an important asymptotic 
 property - an invariance with respect to the 
 non-local dual transformation 
\begin{equation}
   \tpsi(t,\mx) \longleftrightarrow -\varphi(-t,\mx),
   \label{eq:dual_symmetry}
\end{equation}
%
%
\changed{ 
Although not explicitly evident from Eq.~\eqref{eq:action}, this symmetry is revealed once the the field $\psi$ is expressed in terms of $\varphi$ (see also \cite{Janssen1985}) in the following way.
Without the source term, the action can be slightly rewritten using partial integration in time variable
\begin{align*}
  \int dx\, & \biggl[ \partial_t\varphi\left(\partial_t + D_0 \boldnabla^2 - D_0 \tau_0 \right) \tpsi   
  + \frac{u_0 D_0}{2} \tpsi^2 \partial_t\varphi 
  \nonumber \\
  & + 
  \frac{u_0 D_0}{2} (\partial_t \tpsi) \varphi^2
   \biggl].
\end{align*}
By direct algebraic manipulations we can show that this action is indeed invariant with respect to the transformation \eqref{eq:dual_symmetry}.
Let us stress that this symmetry is valid only asymptotically in large time limit, where the term responsible for  the initial condition is discarded.
} 
%
%

 It, however, leads to non-trivial restrictions on 
 critical aspects of the theory.
 
\begin{figure}
\begin{center}
\includegraphics[width = 6cm]{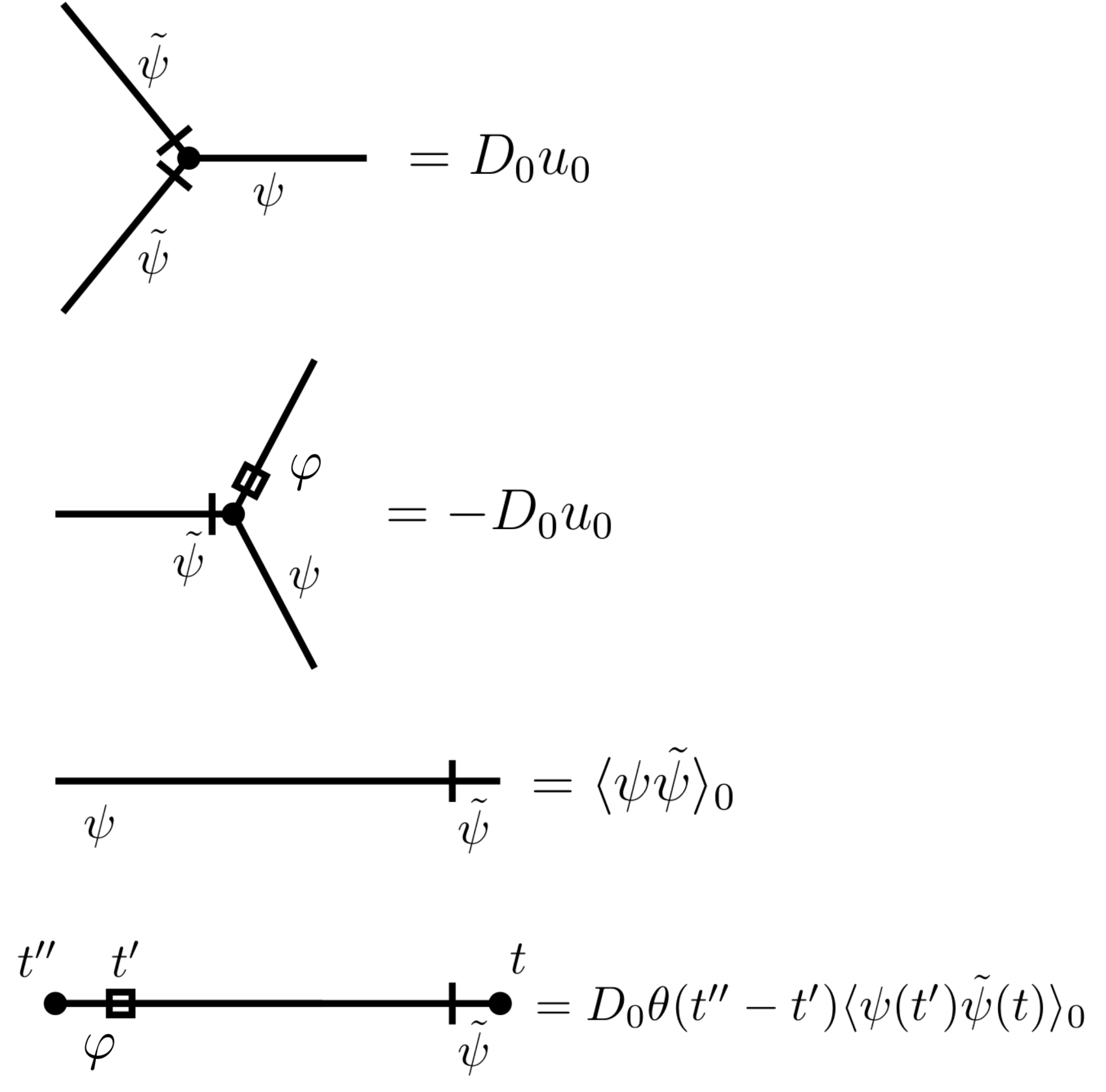}
\caption{Graphical representation of the Feynman rules derived from the field-theoretic action \eqref{eq:action} the dynamical isotropic percolation. Top two elements corresponds to the interaction vertices, whereas the bottom two represents two propagators of the model. Note that in the chosen orientation of the propagators, the time variable should flow from right to left.}
\label{fig:feyn_rules}
\end{center}
\end{figure}

The action \eqref{eq:action} gives rise to a standard representation of perturbation theory, which can be conveniently expressed in the form of Feynman diagrams.
The basic elements are a single propagator and two interaction vertices. 
From the free-part of the action~\eqref{eq:action}
we straightforwardly infer the bare propagator $\langle\psi \tpsi \rangle_0$, which in the a frequency-momentum representation takes the form
\begin{equation}
    \langle \psi \tpsi \rangle_0 (\omega,\mpp)
    = \changed{\frac{1}{-i\omega + D_0(\mpp^2 + \tau_0)}}.
    \label{eq:propagator_omega_mom}
\end{equation}
The nonlinear part of the given action brings about
two interaction vertices $\tpsi^2 \psi$ and $\tpsi\psi\varphi$, respectively. With them, we associate
 vertex factors \cite{Vasilev2004}
 \begin{equation}
    V_{\tpsi \tpsi \psi} = - V_{\tpsi \psi\varphi}
    = D_0 u_0.
    \label{eq:inter_factors}
 \end{equation}
 In graphical terms, all perturbation elements 
 can be found in Fig.~\ref{fig:feyn_rules}.
 
For the present model, it turns out that actual calculations are most easily investigated in the
time-momentum representation $(t,\mpp)$, in which the bare
 propagator~\eqref{eq:propagator_omega_mom} takes the form
\begin{equation}
  \langle \psi  \tpsi \rangle_0 (t,\mpp)
  = \theta (t - t') \exp \left[ -D_0 (\mpp^2 +\tau_0) (t-t') \right].
  \label{eq:propagator_time_mom}
\end{equation}
Note that the introduced perturbation elements effectively permit formulation 
of a second propagator of the type
$\langle \varphi \tpsi \rangle_0$ (see the last element in
Fig.~\ref{fig:feyn_rules} for its graphical depiction). This 
is brought about by combining the integration present in $\varphi \psi \tpsi$ vertex's $\varphi$ field with $\langle \psi \tpsi \rangle_0$ propagator connecting to it. It is convenient to use this propagator in time-momentum representation as
\begin{equation}
    \changed{\langle \varphi(t'') \tpsi(t) \rangle_0 }
    = D_0 
    \int dt'
    \, \theta (t''-t') \langle \psi(t') \tpsi(t) \rangle_0
\end{equation}
with the time variable  $t'$ being integrated over  entire domain (to make the notation 
more compact we will further not explicitly write out the 
 integration 
 $\int dt'$
associated with each such propagator), and for brevity we have suppressed the 
\changed{the remaining spatial (or momentum)}
dependence of the entering fields.

As usual, the construction of perturbation theory in the form of Feynman diagrams is complemented with a conservation law for 
frequency-momentum in each interaction vertex and
 integration over time variable is performed in all internal vertices in a given Feynman diagram.
 Let us remark that the introduced perturbation theory is not unique and there is another permissible ways for defining perturbation elements \cite{Janssen2003}.

The field-theoretic formulation implies, that all the correlation and response functions of the theory can be
 in principle calculated from 
the generating functional
\begin{align} 
  {\cal Z}[A] & = \int [D \Phi] \ \exp\{ -{\cal S}[\Phi] + 
  \Phi A \}, \label{1:eq.Z}
  \\
  \Phi & \equiv \{ \tpsi,\psi \}, \quad
  A \equiv \{ A^{\tpsi}, A^{\psi} \},  
\end{align}
where $[D\Phi]$ is a functional measure and $\Phi A$ is
a shorthand notation for the expression
\begin{equation}
  \int dx \,
  \left[ A^{\tpsi}(x)\tpsi(x) + A^{\psi}(x)\psi(x)
  \right].
  \label{eq:phiA}
\end{equation}
By taking an appropriate number of variational derivatives of ${\cal Z}$ with respect to the source fields ${A} $
we get needed correlation functions. For example, the linear response function is obtained as follows
\begin{align}
  \langle \psi(x) \tpsi (x') \rangle = 
  \frac{\delta^{2} {\cal Z}[ A ]}{\delta A^{\psi} 
  (x) \delta A^{\tpsi}(x')}.
\end{align}
In general, interacting field-theoretic models such as \eqref{1:eq.Z} are not exactly solvable and the only way to make progress is through some perturbation scheme. Here, we utilize a perturbative field-theoretic RG approach. For its effective use, it is  advantageous to work with the effective potential $ \Gamma $ instead, which is defined through a functional Legendre transformation \cite{Vasilev2004,Zinn2007} of   the generating functional~\eqref{1:eq.Z} 
\begin{align}
  \Gamma[\alpha] = \ln {\cal Z}
  [ A 	] - A \alpha, \quad \alpha(x) = 
  \frac{\delta \ln {\cal Z} [ A ]}{\delta A (x)}.
\end{align}
Term $A\alpha$ should be interpreted in an analogous manner to \eqref{eq:phiA}.
It can be shown~\cite{Vasilev2004,Zinn2007}  that the relation between an action functional and its effective potential can be
 compactly written as
\begin{align}
  \Gamma[\alpha] = -{\cal S}[\alpha] + (\text{loop corrections}).
\end{align}
In graphical terms, effective potential $\Gamma$ serves as a generating functional for one-particle irreducible (1PI) Green functions \cite{Vasilev1998,Vasilev2004}. 
 These can be obtained by taking sufficiently
 many variational derivatives of $\Gamma$, i.e.,
\begin{align} 
   \Gamma^{ (N_\Phi, N_{\Phi'}) }(\{ x_i \},\{ x'_j\}) & = \ -
   \frac{\delta^{N_\Phi+N_{\alpha'}} {\cal S}[\alpha]}{\delta \alpha(x_1)  \cdots \delta \alpha'(x_1')\cdots}
   \nonumber\\ 
   & + \text{(loop diagrams)}.  
   \label{1:eq.G}
\end{align}
Here, $N_\Phi$ is a number of all fields appearing in 1PI function, whereas $N_{\Phi'}$ is a number
of all response fields. 
The remaining term in Eq.~\eqref{1:eq.G}
comprises of all 1PI  Feynman
 diagrams, which are such diagrams that remain connected even 
 when any internal line (propagator) is cut \cite{Amit,Vasilev2004}.

{\subsection{Scale invariance}  \label{subsec:scale} } 
 Dynamic models in statistical physics are characterized by two independent correlation lengths, besides the spatial 
  length scale $\xi_{\perp}$ there is also a temporal length scale $\xi_{\parallel}$ \cite{Vasilev2004,Tauber2014}. 
 In the critical region, we expect a system to display strong correlations in time as well in the space domain. Thus the introduced spatiotemporal scales should diverge according to some power-law dependence
\begin{equation}
  \xi_{\perp} \sim |\tau_0|^{-\nu_{\perp}}, \quad \xi_{\parallel} \sim |\tau_0|^{-\nu_{\parallel}},
  \label{eq:anis_scaling}
\end{equation}
where $\tau_0$ was introduced in Eq.~\eqref{eq:basic_eom}.
Usually in statistical physics, exponents $\nu_{\perp}$ and $\nu_{\parallel}$ in Eq.~\eqref{eq:anis_scaling} are not equal.
 Thus, in contrast to relativistic field theories (e.g., quantum electrodynamics), such models
 exhibit so-called strong anisotropic scaling~\cite{Tauber2014}.
 For a given 
 dynamical universality class, it is then convenient to introduce the dynamical exponent $z$ as a ratio
\begin{equation}
   z =  \frac{\nu_{\parallel} }{ \nu_{\perp} }.
   \label{eq:def_exp_z}
\end{equation}  
In the scaling region, the correlation 
 lengths are then simply related as $\xi_{\parallel} \sim \xi_{\perp}^z $.

There are two special values commonly encountered in physics $z=1$ (light-cone spreading)
and $z=2$ (diffusive spreading).
Values $z>2$ correspond to a subdiffusive spreading, whereas $1<z<2$ to a superdiffusive spreading.
 As any other critical exponent $z$ is an universal quantity. 

For the full scaling description of 
%
%
\changed{dIP}
%
%
 process, one additional exponent is needed. Usually, this is the exponent $\beta$, which describes the
 mean particle number in the active phase 
\begin{equation}
  \rho \sim (p_c - p)^\beta \propto (-\tau_0)^\beta.
  \label{eq:def_beta}
\end{equation}
\subsection{Quasi-static limit} \label{subsec:quazistat}
When considering a general universality class, we wish to examine both static and dynamic observables. Regarding the 
%
%
\changed{dIP}
%
%
process, the static properties emerge in the
 asymptotic limit $t \rightarrow \infty$ once all spreading activity of the disease fades out. One way to extract the behavior of observables in this limit is to perform RG analysis for the full dynamical action functional \eqref{eq:action} and 
perform the limit $t\rightarrow \infty$ in the final expressions. However, such an approach entails from the very beginning the use of the full dynamical theory which is a computationally rather demanding task. As observed in \cite{Janssen2005} there exists another more efficient way corresponding to the so-called quasi-static limit arising
in the aforementioned limit $t\rightarrow\infty$. Final
configurations in this limit correspond to the static debris field, description of statistical properties of which are a primary goal.
Since the only interest now consists in Green functions of the debris field at temporal infinity, it is advantageous to switch from agent density field $\psi$ to debris field $\phi$ and its associated response field $\tphi$. We thus perform the following substitutions in the action functional \eqref{eq:action}
\begin{align}
\tpsi (t, \mx) &\rightarrow \tphi (\mx),\\
D_0\int^\infty_0 dt\, \psi(t,\mx) &\rightarrow \phi(\mx).
\end{align}
This yields the quasi-static action functional
%
%
\begin{align}
  \SA_\text{q} = \int d^d x \ \left[ \tphi \left(\boldnabla^2 - \tau_0 \right) \phi + \frac{u_0}{2} \tphi\phi
   \left( \tphi -  \phi \right) \right], \label{eq:action_q}
\end{align}
%
%
where we have explicitly written the required integration
over space variables. Unlike the dynamic action \eqref{eq:action}, no temporal integrations are present in the formula \eqref{eq:action_q}. 
In particular, the bare propagator takes in the momentum representation simple
form $1/(\mk^2+\tau_0)$ and two cubic-like interactions resemble standard models in quantum field theory \cite{kleinert2001}.
The renormalization of this quasi-static theory becomes substantially less demanding. 
In particular, the renormalization factors corresponding to action $\mathcal{S}_q$ coincide with those of the Potts model in the single-state limit \citep{Janssen2005}. The 
corresponding calculations have been performed at the three-loop order many years ago \citep{Bonfirm1981}.
In this way, it is in principle possible to calculate all but one renormalization constant for the 
%
%
\changed{dIP}
%
%
process. In the end, there remains a single renormalization constant that has to be calculated using 
the full dynamical action \eqref{eq:action}. As has been already mentioned in the introcution, the main purpose of the present work is to provide details of such calculation at the three-loop approximation.


{ \section{Renormalization Group Analysis} \label{sec:RG} }

{\subsection{ Canonical dimensions } \label{subsec:can_dim} }
%
%
\changed{ The goal of present analysis is to describe universal properties of the dIP process and its corresponding universality class.}
%
%
In field-theoretic parlance, this is tantamount to determining the scaling behavior of Green functions of 
dynamical fields as functions of space and time in the large spatio-temporal scales. Green functions 
 can be
in the perturbation theory concisely expressed as sums over the Feynman diagrams 
\cite{Vasilev2004,Zinn2007}.
The initial point of the RG analysis consists of a determination of canonical dimensions for all fields and parameters of the model~\cite{Zinn2002,Vasilev2004}.
Since 
%
%
\changed{dIP}
%
%
process is a dynamical model, it is necessary to introduce two independent ca scales (spatial and temporal).
 Therefore, with every quantity $Q$ of the model (field or parameter), we assign two independent canonical dimensions: a frequency dimension $d_{\omega} [Q]$ and
 a momentum dimension $d_k [Q]$. These are determined from the standard condition for the action \eqref{eq:action} to be fully dimensionless, i.e., every term in it has a vanishing canonical dimension with respect to both time and spatial dilatations. The standard normalization
\begin{align}
  d_k [k] & = - d_k [x] = 1, & d_k [\omega] & =  d_k [t] = 0, 
  \label{eq:canon1}  
  \\
  d_{\omega} [k] & = d_{\omega} [x] = 0, & d_{\omega} [\omega] & =  - d_{\omega} [t] = 1 
  \label{eq:canon2}
\end{align}
then fixes the remaining ambiguity \cite{Vasilev2004,Tauber2014}.

Let us also note, that free-theory corresponding to the action
\eqref{eq:action} contains a parabolic differential operator. Usually one is interested in such asymptotic region $\omega\propto k^2$ in which
both terms (proportional to time and Laplace operator, respectively) are relevant. As a consequence both terms $\partial_t \psi$ and $\boldnabla^2 \psi$ have to be taken
 on the equal footing. 
 This reasoning \cite{Vasilev2004} then directly restricts
 the general form of 
 the total canonical dimension $d_Q$ of any quantity $Q$. 
 It is then given as
\begin{equation}
  d [Q] = d_k [Q] + 2 d_{\omega} [Q],
  \label{eq:canon_dim1}
\end{equation} 
where the factor $2$ reflects mentioned asymptotic relation
between frequency and momentum.

The calculation of canonical dimensions then proceeds in a
straightforward manner, and
 relevant values of all fields and parameters for the 
%
%
\changed{dIP}
%
%
 model ~\eqref{eq:action} are summarized in Tab.~\ref{tab:canon_dim}.

The total canonical 
dimension of an 1PI Green function $\Gamma$ takes the form  \cite{Vasilev2004}
\begin{equation}
  d [\Gamma^{(m,n)}] = d_k [\Gamma^{(m,n)}] + 2 d_{\omega}[\Gamma^{(m,n)}].
  \label{eq:canon_dim2}
\end{equation} 
This can be rewritten with the help of Eq.~\eqref{eq:canon_dim1} and Tab.~\ref{tab:canon_dim} 
into a more illuminating 
form
\begin{equation}
  d [\Gamma^{(m,n)}] =  d + 2 - n d[\psi] - m d[\tpsi],
  \label{eq:total_dimension}
\end{equation} 
where $n$ and  $m$ are a number of fields $\psi$ and $\tpsi$ entering given Green function $\Gamma^{(m,n)}$. 

As can be seen in Tab.~\ref{tab:canon_dim}, the model
 is 
%
%
\changed{ logarithmically divergent}
%
%
 at the space dimension $d_c=6$, where the coupling constant $u_0$ becomes dimensionless.
Thus, within dimensional regularization, a small 
parameter $\eps$ is introduced as 
\begin{equation}
  \eps \equiv 
  d_c - d = 6 - d .
  \label{eq:def_eps}
\end{equation}

In the 
%
%
\changed{ logarithmically divergent}
%
%
 theory, the total canonical dimension $d[\Gamma]$ represents the formal degree of UV divergence of the respective 1PI Green function, i.e., 
\begin{equation}  
  \delta_{(m,n)} = d [\Gamma^{(m,n)}]|_{\eps = 0}.
  \label{eq:UV_index} 
\end{equation}
The central step of the following RG procedure constitutes the removal of these superficial UV 
 divergences, which may only appear in those 1PI functions $\Gamma^{(m,n)}$ 
 for which UV index~\eqref{eq:UV_index} 
 $\delta_{(m,n)} $
  has non-negative values.

 Straightforwardly, we determine that for 
%
%
\changed{dIP}
%
%
 model~\eqref{eq:action} the only 1PI functions with UV divergences are  
$\Gamma^{(1,1)}$ and $\Gamma^{(2,1)}$. In principle, three-point Green functions containing one exemplar of all three fields (i.e., $\psi, \tpsi, \phi$) do
have a non-negative index of divergence and should be examined. However, owing to the 
symmetry \eqref{eq:dual_symmetry}, these functions are renormalized with the same factor as functions $\Gamma^{(2,1)}$ and hence we need not calculate them separately.

\begin{table}
\caption{Canonical dimensions of the bare fields and bare parameters for the field-theoretic action~\eqref{eq:action}.}
\begin{ruledtabular}
  \begin{tabular}{c c c c c c}
	Q & $\tpsi, \varphi$ & $\psi$ & $D_0$ & $\tau_0$ & $u_0$ \\
   \noalign{\smallskip}\hline\noalign{\smallskip}
	$d_k [Q]$ & $\frac{d-2}{2}$ & $\frac{d+2}{2}$ & $-2$ & $2$ & $\frac{6-d}{2}$

    \\
	 \noalign{\smallskip}\hline\noalign{\smallskip}
	$d_{\omega} [Q]$ & $0$ & $0$ & $1$ & $0$ & $0$ \\
	  \noalign{\smallskip}\hline\noalign{\smallskip}
	$d [Q]$ &  $\frac{d-2}{2}$ & $\frac{d+2}{2}$ & $0 $ & $2$ & $\frac{6-d}{2}$ 
    
  \end{tabular}
\end{ruledtabular}
\label{tab:canon_dim}
\end{table}

In perturbative calculations, the Green functions are naturally ordered in powers of coupling $u_0$. However, as actual expressions for Feynman diagrams reveal, it is $u_0^2$ that effectively acts as the 
%
%
\changed{ coupling constant}
%
%
 of the theory.
We, therefore, introduce a new charge
\begin{equation}
  g_0 = u^2_0 
  \label{eq:definition_g}
\end{equation}
with total canonical dimension 
$d[g_0] =  \eps$.

Taking into account the considerations above and symmetry \eqref{eq:dual_symmetry}, the renormalized action functional can be presented in the following compact form
\begin{align}
  \mathcal{S}_R &= \int dx\, \biggl[ \tpsi_R(-Z_1\partial_t + Z_2 D  
  \boldnabla^2
  - Z_3D\tau)\psi_R \nonumber \\
  &+
  \frac{ D u \mu^{\eps/2} }{2} Z_4[\tpsi_R^2 \psi_R - 2\tpsi_R \psi_R \varphi_R]
  \biggl],
\label{eq:action_r}
\end{align}
where $\psi_R,\tpsi_R$ are renormalized fields,
 $\mu$ is the arbitrary mass scale in the 
 minimal subtraction (MS) 
 scheme~\cite{Amit,Zinn2002,Vasilev2004} with the dimension $d[\mu]=1$. 
From a technical point of view, this scheme is especially
convenient for the studied problem and we employ it for actual calculations.
  Parameters $Z_i$ $(i=1,2,3,4)$ 
are renormalization constants that have to be determined
 order by order in perturbation theory. Effectively, introduction of these constants into the theory renders
 the model UV finite.
 
 Within the employed MS scheme, the renormalization constants contain only pole parts (in parameter $\eps$) of divergent Feynman diagrams.
 The renormalized counterpart of the debris field $\varphi_R$ follows from its definition \eqref{eq:phi_def} and can be written as
\begin{equation}
  \varphi_R(t,\mx) = D \int\limits_{-\infty}^{t} dt'\, \psi_R(t', \mx).
  \label{eq:phiR_def}
\end{equation}

The fact that the interaction terms in the renormalized action \eqref{eq:action_r} are renormalized by the same renormalization constant $Z_4$ is a direct consequence of 
 the dual symmetry \eqref{eq:dual_symmetry}. 

It was shown previously~\cite{Janssen1981,Janssen2005,Tauber2014} that the model~\eqref{eq:action_r} is multiplicatively renormalizable, and all UV divergences can be absorbed by 
the following prescription
\begin{align}
  \psi & = Z_{\psi} \psi_R, & \tpsi & = Z_{\tpsi} \tpsi_R,
  \label{eq:orig_RG1}
  \\
   \tau_0 & = Z_{\tau} \tau + \tau_c, & u_0 & = \mu^{\eps/2} Z_u u, & D_0 & = Z_D D,
   \label{eq:orig_RG2}
\end{align}
where $Z_Q, Q\in\{\psi,\tpsi,\tau,g,D \}$ are the corresponding renormalization constants that have to be determined perturbatively.
 The term $\tau_c$ denotes the additive renormalization contribution (like a shift of a critical temperature $T_c$ in the theory of critical behavior).
Renormalization of the effective dimensionless charge $g$ directly follows from its definition~\eqref{eq:definition_g}
\begin{equation}
  g_0  = \mu^{\eps} Z_g g, \quad Z_g=Z_u^2.
  \label{eq:renorm_u}
\end{equation}

 Relations between the renormalization constants for fields and parameters and
 renormalization constants $Z_i;i=1,2,3,4$ can be deduced straightforwardly from the renormalized action \eqref{eq:action_r}
\begin{align}
   Z_1 & = Z_{\psi} Z_{\tpsi}, \label{eq:Z1_def}\\
    Z_2 & =  Z_D Z_{\psi} Z_{\tpsi}, \label{eq:Z2_def}\\
    Z_3 & = Z_{\tau} Z_D Z_{\psi} Z_{\tpsi}, \label{eq:Z3_def}\\
   Z_4 & = Z_{u} Z_D Z_{\psi} Z_{\tpsi}^2  =  Z_{u} Z_D^2 Z_{\psi}^2 Z_{\tpsi}. \label{eq:Z4_def}
\end{align}

By direct algebraic manipulations, we readily find
the inverse relations
\begin{align}
  Z_\psi & = Z_1 Z_2^{-1/2},
  &Z_u &= Z_g^{1/2} = Z_4 Z_2^{-3/2}
  \label{eq:rg_relation2a}   
   \\
    Z_{\tpsi} & = Z_2^{1/2},
    &Z_\tau & = Z_3 Z_2^{-1},
    \label{eq:rg_relation2a+}
    \\    
    Z_D & = Z_2 Z_1^{-1}.
   \label{eq:rg_relation2b}          
\end{align}
In actual calculations, we therefore need to analyze renormalization constants $Z_i$, where $i=1,2,3,4$. Moreover, as has been mentioned previously in 
Sec.~\ref{subsec:quazistat}, renormalization constants $Z_2, Z_3, Z_4$ can be calculated within 
 the quasi-static limit and they were calculated analytically to three loop precision in  \cite{Bonfirm1981}. Therefore, it is only the RG constant $Z_1$, whose calculation is missing for the full
 three-loop calculation of the isotropic percolation universality class. 

{\section{Computation of RG Constants} \label{sec:com_rg_const}}
This section is devoted to the crucial steps that allow us a numerical determination of the RG constant $Z_1$. A summary of particular algebraic and technical details on respective Feynman diagrams, such as symmetry factors and their 
pole parts can be found in the supplementary material~\cite{supplement}. 

      To effectively treat ultraviolet
(UV) divergences inherent in Feynman diagrams corresponding to terms in
the renormalized action~\eqref{eq:action_r} 
%
%
\changed{ a}
%
%
renormalization procedure has to be applied \cite{Vasilev2004}. There are various renormalization prescriptions available, each with its own advantages. The general recipe usually consists
of two steps.
First, dimensional regularization has to be employed to formally regularize Feynman diagrams, i.e., to assign a formally convergent expression with each Feynman diagram possessing divergences in the UV region. The role of the regulator is played by the space dimension $d$ \cite{Vasilev2004,Zinn2002}.
Second, as has been already pointed out, we renormalize the model by the MS scheme. In this scheme, UV divergences
manifest themselves in the form of poles in the small 
expansion parameters. For the 
%
%
\changed{dIP}
%
%
, this is given by a deviation from the upper critical dimension $\eps$ introduced in Eq.~\eqref{eq:def_eps}.

It is well-known that in the vicinity of
critical points, large fluctuations on all spatiotemporal scales
dominate the behavior of the system, which in turn results in
the infrared (IR) divergences in the Feynman graphs, i.e., divergences at low frequencies or momenta.
By utilizing the non-trivial relation between UV and IR divergences \cite{Vasilev2004} for a logarithmic theory (see Sec. \ref{subsec:can_dim} for the definition) it is possible to derive RG differential equations, which describe the scaling behavior of the model in the IR region.
As a by-product, an analysis of
these RG equations furnishes an efficient calculational technique for critical exponents.

The renormalization constants can be written in general form as
\begin{equation}
  Z_i (g, \eps) = 1 + \sum\limits_{k=1}^{\infty} g^k \sum\limits_{l=1}^k 
  \frac{c_{i,kl}}{\eps^{l}},  
  \label{eq:general_Z_MS}
\end{equation}
where $c_{i,kl}$ are pure numerical factors (they do not depend on any other model parameters such as temperature variable $\tau$, charge $g$ or diffusion
 constant $D$~\cite{Vasilev2004}). Let us also note that it is convenient to
  absorb a frequently appearing geometric factor into a redefinition of the charge $g$ in the following way
\begin{equation}
  g \frac{\Gamma(1+\eps/2)}{(4\pi)^{3-\eps/2}} \rightarrow g,
  \label{eq:redefinition}
\end{equation}
where $\Gamma(z)$ denotes Euler's gamma function.  
%
%
\changed{ Formally, the rescaling according to the Eq. \eqref{eq:redefinition} amounts to  an use of modified minimal subtraction ($\overline{\text{MS}}$) rather than previously mentioned MS \cite{Collins1984,Zinn2002}}.
%
%
In our calculations, such rescaling is always implied.

In general, counterterms of 1PI functions $\Gamma^{(m,n)}$ have the form of polynomials in external momenta and frequencies and mass parameter 
$\tau_0$, respectively.
 The corresponding degree of divergence for a Green function $\Gamma^{(m,n)}$ is determined by the index 
 $\delta_{(m,n)}$ 
 obtained easily from Eqs.~\eqref{eq:total_dimension} 
 and \eqref{eq:UV_index}.

 To calculate $Z_1$ we only need to consider Green function $\Gamma^{(1,1)}$ (for which $\delta_{(1,1)} = 2$) corresponding to the inverse propagator of the model.
 In addition, we aim to analyze the part of
 $\Gamma^{(1,1)}$ proportional to the external frequency 
 $\Omega$.
Hence, the relevant quantity for the present paper is conveniently 
defined through the following dimensionless expression
\begin{equation}
\Gamma_1  \equiv \partial_{i\Omega} \Gamma^{(1,1)} \bigg|_{\mpp = 0, \Omega=0}. 
\label{eq:Gamma1}
\end{equation}
Let us note that $\Gamma_1$ is normalized in such a way that $\Gamma_1 |_{g_0 = 0} = 1$, and 
 can depend only on the dimensionless ratio $g_0 / \tau_0^{\varepsilon/2}$. 
 Further, the  perturbation expansion for $\Gamma_1$ can
 be arranged in the following way
\begin{equation}
\Gamma_1 (\tau_0, g_0) = 1 + \sum\limits_{n=1} 
(-1)^n
\left(\frac{g_0}{\tau_0^{\varepsilon/2} }
\right)^n
\Gamma_1^{(n)},
\label{eq:LA1}
\end{equation}
where $\Gamma_1^{(n)}$ is a fully dimensionless quantity. 
RG theory \cite{Vasilev2004,Zinn2002} leads to an important relation between renormalized and bare 1PI Green functions
\begin{equation}
  \Gamma^{(m,n)}_R (\ldots; e, \mu)  =
  Z_{\tpsi}^m Z_{\psi}^n 
  \Gamma^{(m,n)} 
  \left( \ldots; e_0(e,\mu) \right),
  \label{eq:RG_relation2}
\end{equation}
where $\ldots$ stands for the common frequency-momentum dependence $\{\omega_i, \mpp_i \}$, $e=\{ D,\tau,g\}$ is a set of all renormalized parameters and $e_0=\{ D_0,\tau_0,g_0\}$ a corresponding set of bare counterparts. Using Eq.~\eqref{eq:RG_relation2} we immediately get for the renormalized analog of the quantity \eqref{eq:Gamma1} useful functional relation
\begin{equation}
\Gamma_{1R} = Z_1 \Gamma_1 \left( \tau Z_{\tau}, \mu^{\varepsilon} g Z_g \right).
\label{eq:LA2}
\end{equation}
Taking into account perturbative expansion \eqref{eq:LA1}, we immediately derive
\begin{align}
\Gamma_{1R} (\tau, \mu, g) & = Z_1 \biggl[ 1 + \sum\limits_{n=1} Z_g^n (-g)^n 
\left( \frac{\mu^2}{\tau } \right)^{n \varepsilon/2} \nonumber\\
& \times Z_{\tau}^{- n \varepsilon/2} \Gamma_1^{(n)} \biggl] .
\label{eq:LA3}
\end{align}

Note that relations such as \eqref{eq:LA1} - \eqref{eq:LA3} hold not only for $\Gamma_1$ but also for analogous expressions created for
other 
relevant 1PI functions:
\begin{align}
\Gamma_2 & = - \frac{1}{2 D_0} \partial_{p}^2 \Gamma^{(1,1)} \bigg|_{\mpp = 0, \omega=0}, 
\label{eq:Gamma2}\\
\Gamma_3 & = - \frac{1}{D_0} \partial_{\tau_0} \Gamma^{(1,1)} \bigg|_{\mpp=0, \omega=0}, 
\label{eq:Gamma3} \\
\Gamma_4 & =  - \frac{1}{u_0 D_0}  \Gamma^{(2,1)} \bigg|_{\mpp=0, \omega=0} .
\label{eq:Gamma4} 
\end{align}
As mentioned earlier in Sec.~\ref{subsec:quazistat}, we do not need to calculate these, as the results are already known.
However, for consistency reasons and verification of our numerical methods,
we still perform such analysis on constants $Z_2$ and $Z_3$ related also to two-point Green function $\Gamma^{(1,1)}$. Our approach yields results that are within a numerical error comparable to analytic expressions obtained from 
the quasi-static limit in three-loops \cite{Bonfirm1981}. For further information, see the supplementary material \citep{supplement}.

The renormalization constants have to be chosen so that the right-hand side 
of Eq. \eqref{eq:LA3}  does not contain poles in $\varepsilon$. It is
 well known from the general renormalization theory \cite{Zinn2002,Vasilev2004} that renormalization constants in the 
%
%
\changed{ $\overline{\text{MS}}$} 
%
%
 scheme do not
  depend on the renormalization mass $\mu$. Hence, an identical requirement can be imposed on the quantity
\begin{equation}
\Gamma_{1R} (\tau = \mu^2, g) = Z_1 \left[ 1 + \sum\limits_{n=1} Z^n (-g)^n 
\Gamma_1^{(n)} \right], 
\end{equation}
where for brevity, we have introduced the abbreviation
\begin{equation}
  Z \equiv 
   Z_g Z_{\tau}^{- \varepsilon/2} .
\end{equation}

The final expressions for RG constant $Z_1$ ($Z_i$) can be additionally validated by their direct insertion into  \eqref{eq:LA3}  and ensuing explicit demonstration 
 that the pole terms containing non-analytic expressions like  $\ln (\mu^2 /\tau)$
 cancel out \cite{Vasilev2004} to a given order.
 Such cancellation provides additional support in favor
 of performed symbolic manipulations. Especially for multi-loop calculations, it is very useful from a technical point of view, because the number of Feynman diagrams is relatively high and it is not feasible to check them manually.

\subsection{Renormalization constant $Z_1$}
\label{subsec:rg_constants}

\begin{table}
\caption{The number of Feynman diagrams for the 
%
%
\changed{dIP}
%
%
model that have to be analyzed 
 to complete the three-loop approximation. }
\begin{ruledtabular}
\begin{tabular}{c  c  c  c}
Diagrams & $1$-loop & $2$-loop & $3$-loop \\
 \noalign{\smallskip}\hline\noalign{\smallskip}
  $\langle \tpsi \psi \rangle$ & $1$ & $6$ & $93$ \\
\end{tabular}
\label{tab:dp_loop}
\end{ruledtabular}
\end{table}
The perturbation expansion for the Green function $\Gamma^{(1,1)}$  consists of the following Feynman diagrams,
 which we 
 classify
 according to the number of loops.
The one-loop contribution includes only one Feynman diagram
\begin{align}
  \raisebox{-2.25ex}{\includegraphics[width=2.cm]{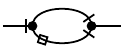}}.
  \label{eq:one_loop_graph}
\end{align}

Further, there are six two-loop diagrams in two distinct topologies
\begin{align}  
    \raisebox{-1ex}{\includegraphics[width=\widthprop]{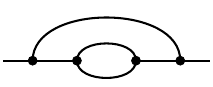}}^{(\times 2)}
  + \quad \raisebox{-3ex}{\includegraphics[width=\widthprop]{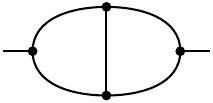}}^{(\times 4)},
\label{eq:two_loop_graphs}
\end{align}
where the indicated number in the parentheses corresponds to the number of distinct diagrams of a given topological kind.
For example, the first topology in Eq. \eqref{eq:two_loop_graphs} contains the following two diagrams
\begin{align}  
    \raisebox{-1.5ex}{\includegraphics[width=\widthprop]{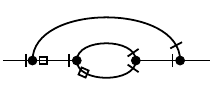}}\ , \quad
   \quad \raisebox{-1.5ex}{\includegraphics[width=\widthprop]{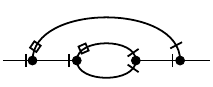}} \ .
\label{eq:two_loop_graphs_top_1}
\end{align}

Note, that one- and two-loop diagrams can be, and have been already calculated analytically previously in works~\cite{Janssen1985,Janssen2005,Amit1976, Chaos2023}.
Finally, the novel result of the present paper is the RG analysis of $93$ Feynman diagrams arising in three-loop order. They can be categorized into the following nine distinct topologies
\begin{align}   
  & \quad
  \raisebox{-3ex}{\includegraphics[width=\widthvertex]{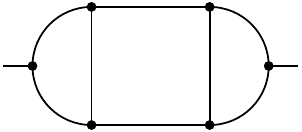}}^{(\times 16)}+ \quad
  \raisebox{-2.ex}{\includegraphics[width=\widthvertex]{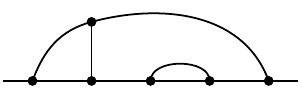}}^{(\times 16)} \nonumber  \\
 &  + \raisebox{-5.ex}{\includegraphics[width=\widthvertex]{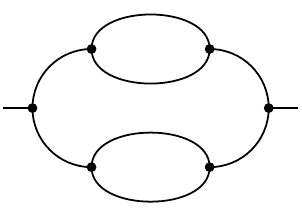}}^{(\times 1)}
 + \quad \raisebox{-3ex} {\includegraphics[width=\widthvertex]{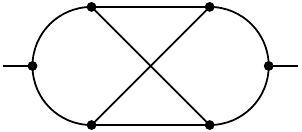}}^{(\times 10)}   \nonumber  \\
 &+  \raisebox{-4.ex}{\includegraphics[width=\widthvertex]{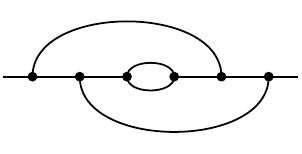}}^{(\times 4)}
  + \quad \raisebox{-3.ex}{\includegraphics[width=\widthvertex]{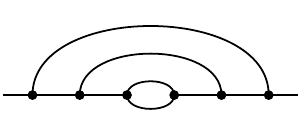}}^{(\times 4)}  \nonumber  \\
  & + \raisebox{-3.ex}{\includegraphics[width=\widthvertex]{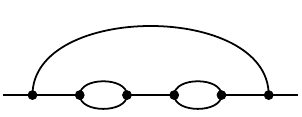}}^{(\times 2)}
   + \quad \raisebox{-4.ex}{\includegraphics[width=\widthvertex]{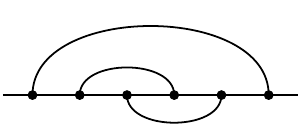}}^{(\times 8)} \nonumber  \\
 & + \raisebox{-2.ex}{\includegraphics[width=\widthvertex]{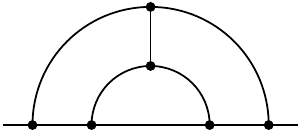}}^{(\times 32)}. 
  \label{eq:three_loop_graphs} 
\end{align}

The actual computations are performed similarly to those of 
directed percolation process (DP) \cite{Adzhemyan2023} or the calculations of model A of critical dynamics \cite{Adzhemyan2018}, respectively.
For our purposes here, two independent implementations of these algorithms have been developed. The first implementation was done in C++ programming language using GiNaC \cite{Bauer2001, Bauer2002} and CUBA \cite{Hahn2005} libraries.
In this approach we encode Feynman diagrams by their Nickel indices with field arguments \cite{Nagle1966, Nickel1977, Batkovich2014},  then construct and symbolically manipulate their integral representations. Feynman parametrization and sector decomposition techniques much like in \cite{Binoth2000} are utilized to identify coefficients of pole terms in each Feynman diagram. These coefficients are then expressed as finite integrals to be later calculated either symbolically using Wolfram Mathematica$^{\circledR}$ \cite{wolfram} (in the case of third-order pole
$\eps^{-3}$) or numerically using CUBA library's Vegas algorithm \cite{Hahn2005}. 
 The second implementation follows our previous work devoted to the three-loop approximation in directed percolation \cite{Adzhemyan2023} employing GraphState \cite{Batkovich2014} and Graphine packages in Python. 
 
 Using the aforementioned two approaches, all particular results in this paper have been independently verified. This supports our belief that the following expressions are justified.  The final expressions that we obtained for the loop contributions of the quantity $\Gamma_1$ read
\begin{align}
    \Gamma_1^{(1)}& = -\frac{3}{4\eps },
    \label{eq:gamma1_1}\\
    \Gamma_1^{(2)}& = \frac{51}{32\eps^2}
    +\frac{1}{\eps}\biggl(\frac{107}{384} - \frac{5}{32} \ln 2  + \frac{9}{64} \ln 3  \biggr) 
    \nonumber \\
    &- 1.25226(2), \label{eq:gamma1_2}\\
    \Gamma_1^{(3)} & = -\frac{527}{128\eps^3}
    -\frac{2.74215(3)}{\eps^2}
    +\frac{7.97518(8)}{\eps},
    \label{eq:gamma1_3}
\end{align}
where $\Gamma_1^{(3)}$ and the finite part of $\Gamma_1^{(2)}$ were computed numerically, and other needed expressions were taken from the previous works \cite{Janssen2005, Janssen1985, Chaos2023}.  Let us stress that the developed programs have been used to verify expressions \eqref{eq:gamma1_1} numerically,  \eqref{eq:gamma1_2} as well as other RG constants as presented in the following section.

From these expressions, we readily get RG constant $Z_1$ to the
three-loop approximation in a straightforward manner as
follows
\begin{widetext}
\begin{align}
Z_{1} = 1 + \frac{3g}{4\eps} +
\frac{g^2}{\eps} \biggl(\frac{102}{64\eps} + \frac{10}{64} \ln 2 - \frac{227}{384}  - \frac{9}{64} \ln 3 \biggr) +
\frac{g^3}{\eps} \biggl( \frac{527}{128\eps^2} - 4.29497 (3) \frac{1}{\eps} + 2.4007(2) \biggr).
	\label{eq:Z1}		
\end{align}
\end{widetext}

 Complete diagram-by-diagram results as well as
  additional relations between diagrams that were used as an additional crosscheck are summarized in supplementary material~\cite{supplement}. 

Furthermore, the information that can be obtained from the cancelation of $\ln (\mu^2 / \tau)$ leads to a relation among the coefficients of the RG constants. We will not give the relations themselves in full, however, the analysis reveals that coefficient of $g^3/\eps^3$ has the same analytic value as
in~\eqref{eq:Z1}. The coefficient of the term $g^3/\eps^2$ ($c_{1,32}$ as introduced in \eqref{eq:general_Z_MS}) has the analytical form 
\begin{equation}
c_{1,32} = -\frac{19135}{4608}+\frac{185}{384}\ln 2 -\frac{111}{256} \ln 3
= -4.294 97,
\end{equation}
where the value was truncated to the last digit. These values are in agreement with those obtained in \eqref{eq:Z1}.
{\subsection{Other RG constants} \label{subsec:other_Zs}}

Results for renormalization constants $Z_2$-$Z_4$ up to three-loop order have been obtained in the past~\citep{Amit1976,Bonfirm1981}.
For the benefit of the reader, we summarize here their values in the 
%
%
\changed{ $\overline{\text{MS}}$} 
%
%
scheme

\begin{widetext}
\begin{align}
Z_2 & =  
1 + \frac{g}{6\eps} + \frac{g^2}{\eps}\biggl(\frac{11}{36\eps}-\frac{37}{432} \biggr) + 
\frac{g^3}{\eps} \biggl( \frac{473}{648 \eps^2} - \frac{1897}{2592 \eps} + \frac{6617}{93312} + \frac{35}{144} - \frac{5}{36} \zeta(3) \biggr),
	\label{eq:Z2}   
   \\ 
Z_3 & = 
1 +\frac{g}{\eps}
+ \frac{g^2}{\eps} \biggl( \frac{9}{4\eps} - \frac{47}{48} \biggr)
+ \frac{g^3}{\eps} \biggl( \frac{6}{\eps^2} - \frac{172}{27\eps} +  \frac{3709}{1296} + \zeta(3)  \biggr),
    \label{eq:Z3}   
   \\ 
Z_4 & =
1 +  \frac{2g}{\eps} 
+ \frac{g^2}{\eps} \biggl( \frac{11}{2\eps} - \frac{59}{24} \biggr)
+ \frac{g^3}{\eps} \biggl( \frac{33}{2\eps^2} - \frac{3643}{216\eps} + \frac{2309}{324} +\frac{11}{3} \zeta(3)  \biggr),
   	\label{eq:Z4}
\end{align}
\end{widetext}
where $\zeta(s) = \sum_{n=1}^\infty 1/n^s$ denotes the Riemann zeta function appearing often in multi-loop calculations \cite{Weinzierl}.
By the subsequent use of inverse relations \eqref{eq:rg_relation2a}
-\eqref{eq:rg_relation2b} we immediately infer
\begin{widetext}
\begin{align}
Z_{\psi} & = 1 + \frac{2g}{3\eps} + 
\frac{g^2}{\eps} \biggl( \frac{25}{18\eps} + \frac{5 \ln 2}{32} - \frac{1895}{3456} - \frac{9 \ln 3}{64}\biggr)
+ \frac{g^3}{\eps} \biggl( \frac{575}{162 \eps^2} - 3.85451(3) \frac{1}{\eps}  + 2.3272(2) \biggr), \\
Z_{D} & = 
1 - \frac{7g}{12\eps} +\frac{g^2}{\eps} \biggl( - \frac{245}{288 \eps} + \frac{1747}{3456} - \frac{5 \ln 2}{32} + \frac{9 \ln 3}{64} \biggr) +
\frac{g^3}{\eps} \biggl(
- \frac{18865}{10368 \eps^2} + 2.77756(3) \frac{1}{\eps} 
- 2.2537(2) \biggr).
\end{align}
\end{widetext} 
 
 The remaining RG constants $Z_{\tpsi}$, $Z_g$, and $Z_{\tau}$ can be obtained from the previous 
 work \citep{Bonfirm1981} alone
 (i.e., Eqs. \eqref{eq:Z2} - \eqref{eq:Z4}) and thus we will refrain from writing them out explicitly here.
 
{\subsection{RG equation and RG functions} \label{subsec:RGeq} }
Once the renormalization constants are known, we can proceed to analyze the asymptotic behavior of the model and calculate the corresponding critical exponents. These exponents are of paramount importance for the quantitative description of the 
%
%
\changed{dIP}
%
%
universality class since they govern the scaling properties of correlation and response functions~\cite{Janssen2005} in
 the critical region $\tau \rightarrow 0$. 

The RG flow can be effectively described by
    the RG differential operator 
\begin{equation}
  \tilde{D}_{\mu} \equiv \mu \partial_{\mu}\bigg|_0, 
\end{equation}
where $\partial_{\mu}|_0$ is the corresponding derivative taken at fixed bare parameters.
Applying it to the fundamental relation between renormalized and bare 1PI Green functions (see Eq.~\eqref{eq:RG_relation2}) yields the basic RG equations~\cite{Amit,Vasilev2004} for renormalized 1PI Green functions 
\begin{equation}  
  \left(\tilde{D}_{\mu} - n \gamma_{\psi} -  m \gamma_{\tpsi}\right)
  \Gamma^{(m,n)}_R ( \ldots; e, \mu) = 0. 
\end{equation}
For the
%
%
\changed{dIP}
%
%
model, similarly to DP, the differential operator $ \tilde{D}_\mu $ 
can be expressed through renormalized variables as
\begin{equation}
  \tilde{D}_{\mu} = \mu \partial_{\mu} + \beta_{g}\partial_{g} - \tau \gamma_{\tau} \partial_{\tau} - D \gamma_D \partial_D,
\end{equation}
where $\beta_g$ is the beta function of the
 coupling $g$, and $\gamma_Q;Q\in\{\psi,\tpsi,D,\tau, g\}$
 are anomalous dimensions for fields and parameters of the model \eqref{eq:action} defined as logarithmic derivatives~\cite{Zinn2002,Vasilev2004} 
\begin{equation}
  \gamma_Q \equiv \mu \partial_{\mu}\bigg|_{0} \ln Z_Q.
  \label{eq:ad_def1}
\end{equation}
These relations can be expressed in a more useful
 form
\begin{equation}
  \gamma_Q  = - \eps \frac{g \partial_g \ln Z_Q}{1 + g\partial_g \ln Z_g},
  \label{eq:ad_def2}
\end{equation}
 by which the anomalous dimension is expressed solely in terms of renormalized variables.
  Since the 
%
%
\changed{dIP}
%
%
  process is a 
%
%
\changed{ model with one coupling constant,}
%
%
   we only have to deal with a single beta function   
\begin{equation}
  \beta_{g} \equiv \mu \partial_{\mu} \bigg|_{0} g. 
  \label{eq:beta_def1}
\end{equation}
Using similar considerations as for anomalous dimensions $\gamma$ in Eq.~\eqref{eq:ad_def2} and
relation \eqref{eq:renorm_u} we can recast the beta function $\beta_g$ into the following form
\begin{align}
\beta_{g} & =  -g 
(\eps + \gamma_g) 
\label{eq:beta_def2} \\
          &= 
          -\eps
          \frac{g}{ 1 + g \partial_g \ln Z_g}. 
\label{eq:beta_def3}
\end{align}
Employing these relations we arrive at anomalous dimensions in the three-loop approximation
\begin{align}
  \gamma_{\psi}  &= -\frac{2g}{3} +  
  \frac{(1895-540 \ln 2 + 486 \ln 3 )}{1728} g^2 \nonumber\\
  & - 6.9816(8) g^3,
  \label{eq:gamma_psi} \\
  \gamma_D &= \frac{7g}{12} + \frac{(-1747 + 
  540 \ln 2  - 486 \ln 3 )}{1728} g^2 \nonumber\\
  & + 6.7611(8) g^3.
     \label{eq:gamma_D}   
\end{align}

As before, only anomalous dimensions that are related to renormalization constant $Z_1$ are explicitly stated.
In fact, we would like to stress that expressions~\eqref{eq:gamma_psi}, \eqref{eq:gamma_D} can be regarded as the most important results of this paper. 
To the best of our knowledge, this is the first time
field-theoretical 
predictions for three-loop corrections are presented for the 
%
%
\changed{dIP}
%
%
model.
Finally, we will state the expressions for the beta function $\beta_g$ and the location of non-trivial stable fixed point. They take in the
three-loop approximation the form \cite{Bonfirm1981}
\begin{align}
 \beta_g &=
 - g \bigg( \eps - \frac{7g}{2} +  \frac{671g^2}{72} 
 - \biggl( \frac{414031}{10368}  + \frac{93}{4} \zeta(3)\biggr) g^3 \bigg),   
 \label{eq:beta_g}\\
   g^* &= 
   \frac{2\eps}{7} + \frac{671\eps^2}{3087}
   + \biggl( \frac{703711}{10890936} - \frac{372}{2401} \zeta(3) \biggr) \eps^3.
\label{eq:g*}
\end{align}
Especially the expression for the nontrivial fixed point $g^*$ -  found from the requirement that $\beta_g$ function \eqref{eq:beta_g}
vanishes and value $\partial \beta_g / \partial g$ has a positive real part - will later enter into the calculation of critical exponents. 
The existence of an IR attractive fixed point \eqref{eq:g*} ensures the scaling behavior of theory~\cite{Amit,Vasilev2004}. 
The asymptotic behavior is then governed by this fixed point. 

Another fixed point of the 
%
%
{\color{red}dIP}
%
%
process corresponds to the trivial (Gaussian) fixed point ($g^* = 0$) and hence to a mean-field solution.
The Gaussian fixed point is IR attractive for $\eps<0$, which according to Eq.~\eqref{eq:def_eps} corresponds to
higher space dimensions $d>d_c$. On the other hand, the non-trivial fixed point is IR stable for $\eps>0$, which corresponds
to physically more relevant space dimensions $d<6$.
 
For a general dynamic model \cite{Vasilev2004}, the critical dimension of the IR relevant quantity $Q$ is given by the general formula
\begin{equation}
  \Delta_Q = d_k [Q] + \Delta_\omega d_\omega [Q] + \gamma^*_Q, 
  \label{eq:critic_dims}
\end{equation} 
where $d_k[Q]$ and $d_{\omega} [Q]$ denote canonical dimensions of quantity $Q$ (see Tab.~\ref{tab:canon_dim}) and $\gamma_Q^*$ is the value of anomalous dimension at the given fixed point. The critical dimension of frequency $\Delta_\omega$ is related to the anomalous
dimension $\gamma_D$ by
\begin{equation}
  \Delta_\omega = 2 - \gamma_D^*,
  \label{eq:critic_dim_freq}
\end{equation} 
where we have assumed the normalization conditions in the form 
\begin{equation}
  \Delta_k = - \Delta_x = 1.
\end{equation}
 The remaining relevant critical dimensions can be expressed through
 critical anomalous dimensions $\gamma_\psi^*$, $\gamma_{\tpsi}^*$ and $\gamma_\tau^*$
\begin{equation}
   \Delta_{\psi} = \frac{d+2}{2} + \gamma_{\psi}^*,\quad \Delta_{\tpsi} = \frac{d-2}{2} + \gamma_{\tpsi}^*, \quad \Delta_\tau = 2 +  \gamma_\tau^*,
\end{equation}
where $\Delta_{\tpsi}$ and $\Delta_\tau$ can be read off from previous result in \citep{Bonfirm1981}. 
 Also, due to a time-reversal symmetry \eqref{eq:dual_symmetry}, only three independent critical exponents are needed to fully describe the 
%
%
\changed{dIP}
%
%
 universality class. 
With $\Delta_\tau$, $\Delta_{\tpsi}$ already at our disposal it is only the
critical exponent
\begin{equation}
    \Delta_\omega = z,
\end{equation}
that has to be found, because it is easy to show that the following relations is satisfied
\begin{equation}
    \Delta_\psi - \Delta_{\tpsi} = \Delta_\omega.
\end{equation}

Here, we thus state the newly acquired critical dimension for the 
%
%
\changed{dIP}
%
%
process 
in three-loop approximation
\begin{align}
\Delta_\omega &= 
2-\frac{\eps}{6} + \biggl(
 \frac{9\ln 3}{392}
- \frac{937}{21168} - \frac{5 \ln 2}{196} \biggr)  \eps^2 \nonumber\\
&+ \biggl( 0.05030(1) \biggr) \eps^3. \label{eq:Delta_omega}
\end{align}
These are to be complemented with previously obtained
static critical indices~\citep{Bonfirm1981}
\begin{align}
\Delta_{\tpsi} &=
 2-\frac{11\eps}{21} - \frac{103\eps^2}{9261}  \nonumber\\
&+  \biggl(-\frac{93619}{2268} + 128 \zeta(3) \biggr) \frac{\eps^3}{7203},
\\
\Delta_{\tau} &= 
2 - \frac{5\eps}{21} - \frac{653\eps^2}{18522} \nonumber \\
&+  \biggl( -\frac{332009}{4084101} + \frac{2848}{7203} \zeta(3) \biggr) \frac{\eps^3}{8}. \label{eq:Delta_tau}
\end{align}
These values were later confirmed and extended to the four-loop precision \cite{gracey2015, pismensky2015, adzhemyan2011}, and later even to the five-loop order \cite{borinsky2021}.

The critical dimensions $\Delta_{\omega}$, $\Delta_{\tpsi}$, and $\Delta_{\tau}$ then represent the full set of critical dimensions for the 
%
%
\changed{dIP}
%
%
process, through which all critical exponents can be expressed in a straightforward manner.
\subsection{Observables} \label{subsec:obs}
\begin{table}
\renewcommand{\arraystretch}{2.2}
\caption{Critical exponents of 
%
%
\changed{dIP}
%
%
universality class. For more exhausting details, we refer the interested reader to the literature~\cite{Janssen2005,Henkel2008}. For the sake of brevity, if permissible, we have suppressed spatial dependence in any quantity.
\changed{ Also the infinity symbol refers to the large time asymptotics $t\rightarrow\infty$. }
}

\begin{ruledtabular}
\begin{tabular}{c | c | c}
Observable & Exponent & Asymptotic relation  \\
\noalign{\smallskip}\hline\noalign{\smallskip}
\multicolumn{3}{c}{Static 
}\\ 
\noalign{\smallskip}\hline\noalign{\smallskip}
Order parameter & $\beta = \frac{\Delta_{\tpsi}}{\Delta_{\tau}}$ & $n(\tau) \sim |\tau|^{\beta}$ \\
\noalign{\smallskip}\hline\noalign{\smallskip}
Survival probability & $\beta' = \beta$ & $P(\tau) \sim |\tau|^{\beta'}$\\
\noalign{\smallskip}\hline\noalign{\smallskip}
 \parbox[c]{3cm}{Correlation length\\spatial}  & $\nu_{\perp} \equiv \nu  = \frac{1}{\Delta_{\tau}}$ & $\xi_{\perp}(\tau) \sim |\tau|^{-\nu_{\perp}}$\\
\noalign{\smallskip}\hline\noalign{\smallskip}
Mean cluster mass & $\gamma = \frac{d - 2\Delta_{\tpsi}}{\Delta_{\tau}}$ & $M(\tau) \sim |\tau|^{-\gamma}$ \\
\noalign{\smallskip}\hline\noalign{\smallskip}
Mean cluster size & $\sigma = \frac{\Delta_{\tau}}{d - \Delta_{\tpsi}}$ & $S(\tau) \sim  |\tau|^{-\sigma}$ \\
\noalign{\smallskip}\hline\noalign{\smallskip}
Pair-correlation fun. & $\eta = 2 - d + 2 \Delta_{\tpsi}$ & $\Gamma(r) \sim r^{-( d - 2 + \eta)} $ \\
\noalign{\smallskip}\hline\noalign{\smallskip}
\multicolumn{3}{c}{Dynamic} \\
\noalign{\smallskip}\hline\noalign{\smallskip}
Mean square radius & $z_s = \frac{2}{ \Delta_{\omega}}$ & $R^2(\tau =0 ,t) \sim t^{z_s}$ \\
\noalign{\smallskip}\hline\noalign{\smallskip}
Number of active sites & $\theta_s = \frac{d - 2\Delta_{\tpsi}}{\Delta_{\omega}} -1$ & $N(\tau = 0, t) \sim t^{\theta_s}$ \\
 \noalign{\smallskip}\hline\noalign{\smallskip}
Survival probability & $\delta_s = \frac{\Delta_{\tpsi}}{\Delta_\omega}$ & $P(\tau = 0, t) \sim t^{-\delta_s}$ \\
\noalign{\smallskip}\hline\noalign{\smallskip}
\parbox[c]{3cm}{Correlation length\\temporal}  & $\nu_{\parallel}  \equiv z \nu  = \frac{\Delta_{\omega}}{\Delta_{\tau}}$ & $\xi_{\parallel}(\tau, \infty) \sim |\tau|^{-\nu_{\parallel}}$ \\   
\end{tabular}
\label{tab:index}
\end{ruledtabular}
\end{table}

Within the 
%
%
\changed{dIP}
%
%
model, we divide observables into two classes: static
 and dynamic observables, respectively. Observables in the former
 class are related to the asymptotic limit $t\rightarrow \infty$. 
Corresponding critical exponents can thus all be calculated knowing two of the critical dimensions, namely $\Delta_{\tpsi}$ and $\Delta_{\tau}$.

Genuine dynamic observables and their respective critical exponents, however, can be only obtained if the critical dimension $\Delta_\omega$ is known. 
%
%
Summary of critical exponents, \changed{ both }static and dynamic,
%
%
and their relations to critical dimensions $\Delta_\omega$, $\Delta_{\tpsi}$ and $\Delta_\tau$ can be found in Table~\ref{tab:index}.
We will further state only the novel results - critical exponents related to dynamcial observables in three-loop order which read

\begin{align}
z_s &= 
1+\frac{\eps}{12} + \biggl(\frac{1231}{108} + 5 \ln 2 - \frac{9}{2} \ln 3 \biggr)\frac{\eps^2}{392} \nonumber \\
& \hspace{1cm} - \, 0.021511(9) \eps^3,\\
\theta_s &= 
\frac{3\eps}{28} + \biggl(\frac{463}{28} + 5 \ln 2 - \frac{9}{2} \ln 3 \biggr)\frac{\eps^2}{392} \nonumber \\
& \hspace{1cm} - \, 0.035614(7) \eps^3,\\
\delta_s&= 1-\frac{5\eps}{28} + \biggl(\frac{167}{252} + 5 \ln 2 - \frac{9}{2} \ln 3 \biggr)\frac{\eps^2}{392} \nonumber \\
& \hspace{1cm} - \, 0.020789 (7) \eps^3, \\
\nu_{||} & = 
1+\frac{\eps}{28} + \biggl(-\frac{25}{252} - 5 \ln 2 + \frac{9}{2} \ln 3 \biggr) \frac{\eps^2}{392} \nonumber \\
& \hspace{1cm} + \, 0.001574(9) \eps^3.
\end{align}

%
%
\begin{table*}
\caption{Dynamical critical exponent estimates using Padé approximants and PBL method. 
The Pad\'{e} approximants are calculated as in \citep{borinsky2021}. The symbol $*$ denotes a value calculated from scaling relations. 
Note, that Pad\'e approximant for  exponent $\theta_s$ was calculated as $\theta_s' = \theta_s + 1$ and subtracted one. In this way we were able to obtain more non-singular approximants for this exponent.
For the benefit of a reader we have also summarized results from the existing literature obtained through simulations \cite{Paul2001, Dammer2004, Zhang2021, Xu2014, munoz1999}.} 
\begin{center}
\begin{ruledtabular}
\renewcommand{\arraystretch}{1.5}
\begin{tabular}{r l l l l l l}
$d$ & 
& $z_s$ & $z$ & $\theta_s$ & $\delta_s$ & $\nu_{\parallel}$\\
\hline
$6$ & 
& $1$ & $2$ & $0$ & $1$ & $1$ \\
\hline
    & Pad\'{e} 
    & 
     \changed{$1.11 (2)$}
    & \changed{$1.81(4)$}  & \changed{$0.14(4)$}  & $0.819$ & $1.041(5)$ \\
    & PBL & 
    $1.099(7)^{*}$ & $1.820(12)$ & $0.130(3)^{*}$ & $0.809(2)^{*}$ & $1.045(5)^*$\\
$5$ & \cite{Paul2001} & 
    $1.104(4)^{*}$ & $1.812(6)$ & $0.149(12)^{*}$ & $0.805(6)^{*}$ &  \\
    & \cite{Dammer2004} &
    $1.102(6)^{*}$ & $1.815(10)$ & $0.134(10)$ & $0.806(12)$ &  \\
    & \cite{Zhang2021} & 
    $1.1027(5)^{*}$ & $1.8137(16)$ & $0.1314(18)^{*}$ & $0.8127(11)^{*}$ & $1.041(6)^{*}$ \\
\hline
    & Pad\'{e} &
    \changed{$1.22(5)$} & 
    \changed{$1.6(3)$}
    & $0.29(10)$ & $0.634$ & $1.089$ \\
    & PBL & 
    $1.22(5)^{*}$ & \changed{$1.64(7)$} & $0.28(2)^{*}$ & $0.578(11)^{*}$ & $1.13(4)^{*}$ \\
$4$ & \citep{Paul2001} & 
    $1.245(4)^{*}$ & $1.607(5)$ & $0.302(10)^{*}$ & $0.593(5)^{*}$ &  \\
    & \cite{Dammer2004} & 
    $1.246(7)^{*}$ & $1.605(9)$ & $0.30(1)$ & $0.595(8)$ & \\
    & \citep{Zhang2021} & 
    $1.2467(2)^{*}$  & $1.6042(5)$ & $0.3023(10)^{*}$ & $0.5955(5)^{*}$ & $1.098(4)^{*}$ \\
\hline
    & Pad\'{e} & 
    $1.35(17)$ & $1.48(19)$ & \changed{$0.5(3)$}  & $0.445$ & $1.152$ \\
    & PBL & 
    $1.35(17)^{*}$ & $1.48(19)$ & $0.39(7)^{*}$ & $0.32(2)^{*}$ & $1.32(12)^{*}$ \\
$3$ & \cite{Dammer2004} & 
    $1.454(5)^{*}$ & $1.375(5)$ & $0.488(7)$ & $0.346(6)$ & \\
    & \citep{Xu2014} & 
    $1.4540(3)^{*}$ & $1.3755(3)$ & $0.4874(4)^{*}$ & $0.3468(1)^{*}$ &  $1.2052(16)^{*}$\\
    & \citep{munoz1999} & 
    $1.497^{*}$ &  $1.336^{*}$ & $0.5321^{*}$ & $0.3567^{*}$ & $1.169$ \\
\hline
    & Pad\'{e} & 
    \changed{$1.5(4)$} & \changed{$1.3(4)$} & \changed{$0.8(8)$}  & $0.251$ &  $1.236$ \\
    & PBL & 
    \changed{$1.5(4)^{*}$} & \changed{$1.3(4)$} & 
    \changed{$0.2(2)$}
    & $0.14(9)^{*}$ & \changed{$1.5(3)^{*}$} \\
$2$ & \citep{Zhang2021} & 
    $1.76871(3)^{*}$ & $1.13077(2)$ & $0.58447(3)^{*}$ & $0.092120(2)^{*}$ & $1.50769(2)^{*}$\\
    & \citep{munoz1999} & 
    $1.771^{*}$ &  $1.1295^{*}$ & $0.586^{*}$  & $0.092^{*}$  & $1.506$
\end{tabular}
\end{ruledtabular}
\end{center}
\label{tab:exp_results}
\end{table*}
 The final results of critical exponents are expressed in the form of asymptotic series in formally small expansion parameter $\eps = 6 - d$.  
To obtain final values in lower space dimensions $d=2$ ($\eps_{phys} = 4$) up to $d=5$ ($\eps_{phys}=1$) we have employed Pad\'{e} approximants 
(see Table~\ref{tab:exp_results}), where the analysis was performed following the strategy presented previously in \cite{borinsky2021,Adzhemyan2019}.

The complicating factor in present analysis is that the third-order approximants, 
in most cases, contain poles in the interval $\eps \in [0,2\eps_{phys}]$. 
Such approximants cannot give reliable estimates because of the unphysical pole contribution, and are removed from further analysis. The error bars are then estimated from the remaining approximants. Note, however, that in the case of exponents $\delta_s$ and $\nu_{\parallel}$ not enough approximants remain to determine the error bar and these values are stated without it. 

We also provide the estimates for relevant critical exponents using the resummation method built on Pad\'{e} approximants augmented with the Borel-Leroy transformation (PBL) \cite{Adzhemyan2019}. 
Here, it was only the exponent $z$ that was obtained and other values of other exponents were calculated
from the scaling relations, with 
the static exponents $\beta$, $\nu_{\perp}$ taken from the best estimation in five loops \cite{borinsky2021}. The obtained values are also summarized in Table \ref{tab:exp_results}.
Conformal mapping (KP17 \cite{borinsky2021}) can also be performed, however, the estimations are much worse compared to Pad\'e or PBL.
%
%
 This is to be expected as for the conformal mapping to give convergence better than Pad\'e or PBL, more terms are required (in case of $\phi^3$ theory \cite{borinsky2021} it is only in five-loop order that conformal mapping overcomes Pad\'{e} and PBL significantly).
Therefore, in the case of present three-loop study, this method is not further used, neither we explicitly state the corresponding results.
%
%

%
%
 The Table \ref{tab:exp_results} also contains the values for dynamical exponents obtained by other methods for comparison. These are, in existing literature, limited to Monte Carlo simulations \cite{Zhang2021,Dammer2004,Paul2001,Xu2014,munoz1999}. To authors knowledge, no reasonable experimental data on dynamical properties of isotropic percolation are available and hence, not stated. 
%
%

We conclude that estimates given by both resumation techniques are in good agreement with simulations for dimensions $d=4$ ($\eps_{phys}=2$) and $d=5$ ($\eps_{phys}=1$).
In lower dimensions, namely $d=2$ ($\eps_{phys}=4$) and $d=3$ ($\eps_{phys}=3$), the discrepancy is larger. Nonetheless, the results provide a rough estimation in the three-loop perturbation approach.

In order to obtain even better estimates for critical exponents, especially for these spatial dimensions, it is necessary to perform similar analysis in higher orders of perturbation theory. 
Such analysis, however, goes beyond the scope of the present paper and is deferred to future work.


{\section{Conclusion}	\label{sec:conclusion} }

In this paper, we have analyzed dynamic isotropic percolation
process utilizing field-theoretic perturbative renormalization
group.
We have formulated the problem in terms of the functional integral since this is 
 arguably the most economical approach to perturbative calculations of universal quantities. Within this framework, we
 have presented major points of the
renormalization analysis. We have restricted ourselves mainly to the calculation computing only the dynamic exponents related to $z = \Delta_\omega$ as
the other necessary critical exponents have been already
obtained by different means. 
To the best of our knowledge, this is the first time
the dynamic exponent $z$ for the dynamic isotropic percolation process has been calculated in the three-loop order. Thus, we have completed the three-loop calculation of the corresponding universality class.

We give perturbative $\eps$-expansion for the exponent $z$ in the form of an asymptotic series. Our results are in agreement with the two-loop analytic calculations.

 In the calculation of the RG constants, we have identified three distinguished properties, which have helped us simplify 
the demanding technical part related to the analytical treatment of
three-loop diagrams.
 Moreover, we have used them for independent crosschecks to verify the obtained 
 expressions. Their validity is also supported by cancelation of non-analytic logarithmic terms in renormalized 1PI functions. Further, on top of novel results stemming from renormalization constant 
 $Z_1$ we have also performed computation of constants $Z_2$, $Z_3$. Though these are already known up to three-loop order this analysis served for purposes of verification of our approach as these agree with existing analytical and numerical results. Analysis of $Z_4$ constant was not performed
  (although, in principle, possible) due to excessive computational load if attempted from the full dynamical action.
 
  In the space dimensions $d = 5$ ($\eps_{phys} = 1$) up to $d = 2$ ($\eps_{phys}=4$), we have carried out
 the resummation technique using Pad\'{e} approximants and PBL method. The exponent values in dimensions $d=5$ and $d=4$ display good agreement with the simulation results \cite{Zhang2021,Dammer2004,Paul2001}. However, for the the physically relevant spatial dimensions $d=3$ and $d=2$ the used resummation methods do not provide a satisfactory quantitative estimate.
%
%
\changed{ Whether this shortfall can be mended by obtaining higher-order results remains an open question, and is deferred to future work.}
%
%

In our view, this paper presents a further step towards enhancing multi-loop calculations in the area of non-equilibrium models. We believe that it may provide additional support and also open new possibilities to solve other more involved dynamic models and thus provide
more insight into their perturbation structure. 
\begin{acknowledgments}
The work was supported by VEGA grant No. 1/0297/25 of the Ministry of Education, Science, 
Research and Sport of the Slovak Republic and project VVGS-2023-2567 of the internal grant scheme of Faculty of Science, \v{S}af\'{a}rik University in Ko\v{s}ice.

The authors have greatly benefited from mutual collaborations and many valuable discussions with Loran~Ts.~Adzhemyan.
\end{acknowledgments}
\appendix
{\section{Construction of integrand for the two-point Green functions proportional to external frequency}	\label{sec:appendixa} }

As has been mentioned in Sec. \ref{sec:DIP} the Feynman diagram technique applied to the action functional \eqref{eq:action} gives rise to two interaction vertices with respective vertex factors $V_{\psi \tpsi \tpsi}$, $V_{\varphi \psi \tpsi}$, and two propagators $\langle \psi \tpsi \rangle_0$ and $\langle \varphi \tpsi \rangle_0$, respectively.

To explain the following, it is essential to start with propagators in frequency-momentum representation. These take the form
\begin{align}
\left\langle \psi \tpsi\right\rangle_0(\mpp, \omega)  &= \frac{1}{-i\omega + D_0(\mpp^2 + \tau_0)}, \\
 \left\langle \varphi \tpsi \right\rangle_0 (\mpp, \omega) & =
 \frac{1}{-i\omega} \left\langle \psi \tpsi\right\rangle_0(\mpp, \omega) .
\end{align}

\begin{figure*}
\begin{center}
\includegraphics[width = \textwidth]{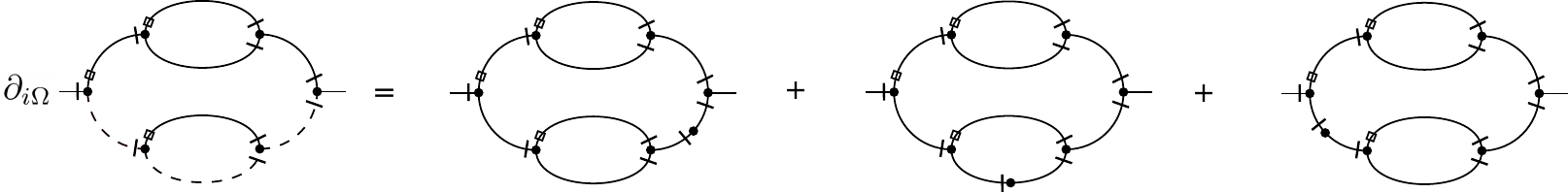}
\caption{Upon action of the partial derivative 
$\partial_{i\Omega}$ the given Feynman diagram breaks up into the sum of diagrams, each of which contains one of the $\langle \psi \tpsi \rangle_0$ propagators carrying external frequency split as shown in Eq. \eqref{appendixAeq:prop_splitting}. We chose the dashed line to represent propagators carrying external momentum/frequency.}
\label{appendixAfig:unit_vertex}
\end{center}
\end{figure*}

We are mainly interested in part of the two-point Feynman diagrams ($\Gamma^{(1,1)}$) proportional to the external frequency $\Omega$.
For this purpose, it is convenient to look for
\begin{equation}
\partial_{i\Omega} \Gamma^{(1,1)},
\end{equation}
as every propagator carrying external frequency does indeed carry the expression $i\Omega$ rather
than just frequency $\Omega$ alone. Also, the results ought not to depend on the choice of momentum flow, it is then convenient to always choose such flow in which the external frequency (and momentum for that matter) are carried by the simpler (with respect
to a functional form) $\langle \psi \tpsi \rangle_0$ propagators.
Applying the derivative on such propagator yields
\begin{align}
\partial_{i\Omega} \langle \psi \tpsi \rangle_0 = 
[-i(\Omega + \omega) + D_0((\mpp +\mk)^2 + \tau_0)]^{-2}.
  \label{appendixAeq:prop_splitting}
\end{align}
Returning to the convenient time-momentum representation we observe that the derivative with respect to external frequency produced a unit vertex on the propagator it acted on thus splitting it in two. In diagrammatic language, it
can be schematically written as
\begin{equation}
  \partial_{i\Omega}
  \raisebox{-4.5pt}{\includegraphics[width = 1.5cm]{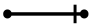}}   
  =
  \raisebox{-4.5pt}{\includegraphics[width = 2.5cm]{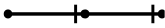}  }
  .
  \label{eq:deriv_prop}
\end{equation}
In the rest of the diagram (corresponding integral), we are now allowed to set both external frequency and external momentum to zero ($\Omega = 0$, $\mpp = 0$).

Each diagram after the application of $\partial_{i\Omega}$ thus breaks up into a sum of new diagrams, each of which has exactly one propagator carrying external frequency split by the unit vertex insertion. 
An example of this can be seen in Fig. \ref{appendixAfig:unit_vertex}.

{\section{Notation differences and conventions in existing works}	\label{sec:appendixb} }
We would like to note, that 
%
%
\changed{dIP}
%
%
universality class has been analysed before by means of both RG and computer simulation methods. Most of these are mentioned and results of some of these works are directly used in present paper. The conventions regarding the notation, however, are almost always article-specific.

Hereby, we would like to state at least some of other notation used, namely in papers \cite{Bonfirm1981} and \cite{Zhang2021, Paul2001} as these are used as three-loop analytical results regarding static aspects of the model, and Monte Carlo simulations with which we compare the final estimates respectively.

Whereas we defined all the critical exponents in terms of critical dimensions $\Delta_\omega$ and $\Delta_{\tpsi}$, $\Delta_\tau$ (see Eq. \eqref{eq:Delta_omega}-\eqref{eq:Delta_tau}), with the latter two being related to statics,
in the paper \cite{Bonfirm1981} the value of anomalous dimensions are presented as follows
\begin{equation}
\eta \equiv \gamma_2^{*},  \hspace{1cm} \nu^{-1} - 2 + \eta \equiv \gamma_3^{*},
\end{equation}
where $\gamma_i, \ i \in \{2,3\}$ is defined as in \eqref{eq:ad_def1} but for renormalization constants $Z_2$ and $Z_3$.
In \cite{Bonfirm1981} it is then through $\eta$ and $\nu$ that other critical exponents are expressed.

The critical exponents can also be found yet in another formulation in works \cite{Zhang2021, Paul2001} on lattice simulations. Here the critical exponents are associated to the geometric properties of the lattice cluster and can be expressed in critical dimensions as
\begin{align}
\textrm{short-path exponent: } & d_{min} = \Delta_{\omega}, \\
\textrm{thermal exponent: } & y_t = \Delta_{\tau}, \\
\textrm{fractal dimension: } & d_f = y_h = d - \Delta_{\tpsi}.
\end{align}

In the end, the specific choice always depends on the problem being solved, and on personal preference. To fully determine the 
%
%
\changed{dIP}
%
%
universality class, however, only three critical exponents are needed. These being one dynamic and two static exponents, regardless of particular choice.

\bibliographystyle{apsrev} 
\bibliography{mybib}

\end{document}


\title{Supplementary Material for\\ ``Field-theoretic Analysis of Dynamic Isotropic Percolation: Three-loop Approximation''}

\author{Michal Hnati\v{c}}%
\email{hnatic@saske.sk}
\affiliation{Bogolyubov Laboratory of Theoretical Physics, Joint Institute for Nuclear Research, 141980 Dubna, Russian Federation,}
\affiliation{Institute of Experimental Physics SAS, Watsonova 47, 040 14 Ko\v{s}ice, Slovakia}
\affiliation{Faculty of Science, \v{S}af\'{a}rik University, Moyzesova 16, 040 01 Ko\v{s}ice, Slovakia}

\author{Matej Kecer}
\email{matej.kecer@student.upjs.sk}
\affiliation{Faculty of Science, Pavol Jozef \v{S}af\'{a}rik University, Moyzesova 16, 040 01 Ko\v{s}ice, Slovakia}

\author{Mikhail V. Kompaniets}
\email{m.kompaniets@spbu.ru}
\affiliation{Bogolyubov Laboratory of Theoretical Physics, Joint Institute for Nuclear Research, 141980 Dubna, Russian Federation}
\affiliation{Sankt Petersburg State University, St. Petersburg 199034, Russian Federation}

\author{Tom\v{a}\v{s} Lu\v{c}ivjansk\'{y}}
\email{tomas.lucivjansky@upjs.sk}
\affiliation{Faculty of Science, Pavol Jozef \v{S}af\'{a}rik University, Moyzesova 16, 040 01 Ko\v{s}ice, Slovakia}

\author{Luk\'{a}\v{s} Mi\v{z}i\v{s}in}
\email{mizisin@theor.jinr.ru}
\affiliation{Bogolyubov Laboratory of Theoretical Physics, Joint Institute for Nuclear Research, 141980 Dubna, Russian Federation}

\author{Yurii \,G.\,Molotkov}
\email{molotkov@theor.jinr.ru}
\affiliation{Bogolyubov Laboratory of Theoretical Physics, Joint Institute for Nuclear Research, 141980 Dubna, Russian Federation}

\date{\today}
\maketitle

{\section{Feynman diagrams contributing to $Z_1$} \label{appendix:diagram}}

To analyze Feynman diagrams in high-orders of perturbation theory, it is necessary to have at our disposal an effective
way for their mathematical representation. This is especially important for any symbolic manipulation, as typical multi-loop
calculations are rather tedious and lengthy.
 In our work, we have enumerated Feynman diagrams according to a convenient and efficient scheme
  based on Nickel index with field arguments \citep{Nagle1966,Nickel1977,Batkovich2014}, representing each Feynman diagram unambiguously.
 Details and examples can be found in the existing literature, e.g., \citep{Batkovich2014}.

We group three-loop two-point Feynman diagrams with the same topology and denote them by respective diagram number (e.g. diagram $T_1(1)$ stands for topology $T_1$, diagram No. $1$). The graphical representation of all topologies is depicted in Tab. \ref{appendixtab:topologies}.

Results for all the relevant Feynman diagrams are summarized in Tab.~\ref{appendixtab:T1} - Tab.~\ref{appendixtab:T9}.
Graphical representation for each diagram can be obtained directly from its Nickel index. Each table states the diagram number, its Nickel index representation, symmetry factor (S.F.), and coefficients of poles in its divergent part.
Sum of all the diagrams in respective topology is stated under each table.
Note, that the coefficient of the highest pole term ($1/\eps^3$) was calculated symbolically using Wolfram Mathematica software \cite{wolfram}. Other pole coefficients were computed numerically using Cuba packages \cite{Hahn2005} Vegas algorithm. Hence, numerical errors are always stated in parentheses. All the decimal numbers stated have been truncated.

Note, that compared to the main article, here in the supplement material, all pole terms are expressed using the parameter $\eps = (6-d)/2$, which was our working definition. In the main article, the results are written using, in the literature more common, small parameter $\varepsilon = 2\eps = 6-d$.
The divergent part of general Feynman diagram in DyIP in three-loop approximation has the form 
\begin{equation}
T_j (i) = \frac{\Gamma(1+3\eps)}{(4\pi)^{9-3\eps}} \tau_{0}^{-3\eps} \frac{u_0^6}{3 \eps} \left(\frac{a_3}{\eps^{2}} + \frac{a_2}{\eps} + a_1 \right). \label{suppeq:contribution}
\end{equation}
The results for particular diagrams will be written following this form, stating the coefficients $a_3, a_2$ and $a_1$.

\begin{table}[!ht]
\centering
\begin{tabular}{c @{\hspace{1cm}} c @{\hspace{1cm}}  c}
Topology $T_1$ & Topology $T_2$ & Topology $T_3$ \\
\raisebox{-5.8ex}{\includegraphics[width=4.cm]{Figures/T1.pdf}} &  
\raisebox{0ex}{\includegraphics[width=4.cm]{Figures/T2.pdf}} &
\raisebox{-9.8ex}{\includegraphics[width=4.cm]{Figures/T3.pdf}} \\
& & \\
Topology $T_4$ & Topology $T_5$ & Topology $T_6$ \\
\raisebox{1.2ex}{\includegraphics[width=4.cm]{Figures/T4.pdf}} &
\raisebox{0.ex}{\includegraphics[width=4.cm]{Figures/T5.pdf}} &
\raisebox{5.3ex}{\includegraphics[width=4.cm]{Figures/T6.pdf}} \\
& & \\
Topology $T_7$ & Topology $T_8$ & Topology $T_9$ \\
\raisebox{0.9ex}{\includegraphics[width=4.cm]{Figures/T7.pdf}} & 
\raisebox{-1.ex}{\includegraphics[width=4.cm]{Figures/T8.pdf}}  &
\raisebox{2.2ex}{\includegraphics[width=4.cm]{Figures/T9.pdf}}
\end{tabular}
\caption{Graphical representation of all diagram topologies in the three-loop approximation.}
\label{appendixtab:topologies}
\end{table}


\renewcommand\arraystretch{1.5}
\begin{center}
\begin{longtable}{|c|c|c|c|c|c|}
\caption{Feynman diagrams belonging to the topology $T_{1}$.} \label{appendixtab:T1} \\

\hline \multicolumn{1}{|c|}{ No.} & \multicolumn{1}{|c|}{ 
%
%
Nickel index \added{with field arguments}
} & \multicolumn{1}{|c|}{ S.F. } & \multicolumn{1}{|c|}{$a_3$} & \multicolumn{1}{|c|}{$a_2$} & \multicolumn{1}{|c|}{$a_1$}
\\ \hline 
\endfirsthead

\multicolumn{4}{|c|}%
{{\bfseries \tablename\ \thetable{} -- continued from previous page}} \\
\hline \multicolumn{1}{|c|}{ No. } &
\multicolumn{1}{|c|}{ 
%
%
Nickel index \added{with field arguments}
} &
\multicolumn{1}{|c|}{ S.F. } & \multicolumn{1}{|c|}{$a_3$} & \multicolumn{1}{|c|}{$a_2$} & \multicolumn{1}{|c|}{$a_1$} \\ \hline 
\endhead

\hline \multicolumn{6}{|c|}{{Continued on next page}} \\ \hline
\endfoot

\hline \hline
\endlastfoot
1 & $e12|23|4|45|5|e|:0\tpsi\_\phi\tpsi\_\psi\tpsi|\tpsi\psi\_\psi\tpsi|\phi\tpsi|\psi\tpsi\_\phi\tpsi|\psi\tpsi|0\psi|$ & $1$ & $-27/256$ & $-0.464487(6)$ & $1.267215(10)$\\ \hline
2 & $e12|23|4|45|5|e|:0\tpsi\_\phi\tpsi\_\psi\tpsi|\psi\tpsi\_\phi\tpsi|\psi\tpsi|\phi\tpsi\_\psi\tpsi|\psi\tpsi|0\psi|$ & $1$ & $-15/1024$ & $0.0021089(7)$ & $0.055210(1)$\\ \hline
3 & $e12|23|4|45|5|e|:0\tpsi\_\phi\tpsi\_\psi\tpsi|\tpsi\phi\_\psi\tpsi|\psi\tpsi|\tpsi\psi\_\psi\tpsi|\phi\tpsi|0\psi|$ & $1$ & $-117/1024$ & $-0.298115(5)$ & $1.147283(13)$\\ \hline
4 & $e12|23|4|45|5|e|:0\tpsi\_\phi\tpsi\_\psi\tpsi|\phi\tpsi\_\psi\tpsi|\psi\tpsi|\tpsi\phi\_\psi\tpsi|\psi\tpsi|0\psi|$ & $1$ & $-3/1024$ & $0.0031764(1)$ & $-0.0014845(9)$\\ \hline
5 & $e12|23|4|45|5|e|:0\tpsi\_\phi\tpsi\_\psi\tpsi|\tpsi\psi\_\psi\tpsi|\phi\tpsi|\phi\tpsi\_\psi\tpsi|\psi\tpsi|0\psi|$ & $1$ & $-63/1024$ & $-0.101834(2)$ & $0.285054(9)$\\ \hline
6 & $e12|23|4|45|5|e|:0\tpsi\_\phi\tpsi\_\psi\tpsi|\tpsi\phi\_\psi\tpsi|\psi\tpsi|\tpsi\phi\_\psi\tpsi|\psi\tpsi|0\psi|$ & $1$ & $-75/1024$ & $-0.0451642(12)$ & $0.269435(9)$\\ \hline
7 & $e12|23|4|45|5|e|:0\tpsi\_\phi\tpsi\_\psi\tpsi|\tpsi\psi\_\psi\tpsi|\phi\tpsi|\tpsi\psi\_\psi\tpsi|\phi\tpsi|0\psi|$ & $1$ & $-45/512$ & $-0.231998(3)$ & $0.873449(9)$\\ \hline
8 & $e12|23|4|45|5|e|:0\tpsi\_\phi\tpsi\_\psi\tpsi|\phi\tpsi\_\psi\tpsi|\psi\tpsi|\phi\tpsi\_\psi\tpsi|\psi\tpsi|0\psi|$ & $1$ & $-3/1024$ & $-0.0136140(2)$ & $0.0204983(8)$\\ \hline
9 & $e12|23|4|45|5|e|:0\tpsi\_\phi\tpsi\_\psi\tpsi|\tpsi\phi\_\psi\tpsi|\psi\tpsi|\phi\tpsi\_\psi\tpsi|\psi\tpsi|0\psi|$ & $1$ & $-27/1024$ & $0.0028969(9)$ & $0.148183(2)$\\ \hline
10 & $e12|23|4|45|5|e|:0\tpsi\_\phi\tpsi\_\psi\tpsi|\psi\tpsi\_\phi\tpsi|\psi\tpsi|\psi\tpsi\_\phi\tpsi|\psi\tpsi|0\psi|$ & $1$ & $-3/64$ & $-0.0070383(14)$ & $0.166956(9)$\\ \hline
11 & $e12|23|4|45|5|e|:0\tpsi\_\phi\tpsi\_\psi\tpsi|\phi\tpsi\_\psi\tpsi|\psi\tpsi|\tpsi\psi\_\psi\tpsi|\phi\tpsi|0\psi|$ & $1$ & $-9/1024$ & $0.0043616(6)$ & $-0.003798(3)$\\ \hline
12 & $e12|23|4|45|5|e|:0\tpsi\_\phi\tpsi\_\psi\tpsi|\tpsi\psi\_\psi\tpsi|\phi\tpsi|\tpsi\phi\_\psi\tpsi|\psi\tpsi|0\psi|$ & $1$ & $-27/1024$ & $-0.0661174(10)$ & $0.273833(9)$\\ \hline
13 & $e12|23|4|45|5|e|:0\tpsi\_\phi\tpsi\_\psi\tpsi|\phi\tpsi\_\psi\tpsi|\psi\tpsi|\psi\tpsi\_\phi\tpsi|\psi\tpsi|0\psi|$ & $1$ & $-9/1024$ & $-0.0266206(5)$ & $0.026769(2)$\\ \hline
14 & $e12|23|4|45|5|e|:0\tpsi\_\phi\tpsi\_\psi\tpsi|\tpsi\phi\_\psi\tpsi|\psi\tpsi|\psi\tpsi\_\phi\tpsi|\psi\tpsi|0\psi|$ & $1$ & $-93/1024$ & $0.0025489(16)$ & $0.486570(8)$\\ \hline
15 & $e12|23|4|45|5|e|:0\tpsi\_\phi\tpsi\_\psi\tpsi|\psi\tpsi\_\phi\tpsi|\psi\tpsi|\tpsi\psi\_\psi\tpsi|\phi\tpsi|0\psi|$ & $1$ & $-27/512$ & $-0.106195(3)$ & $0.288852(7)$\\ \hline
16 & $e12|23|4|45|5|e|:0\tpsi\_\phi\tpsi\_\psi\tpsi|\psi\tpsi\_\phi\tpsi|\psi\tpsi|\tpsi\phi\_\psi\tpsi|\psi\tpsi|0\psi|$ & $1$ & $-27/1024$ & $-0.0270261(5)$ & $0.067221(4)$\\ \hline

\end{longtable}
\end{center}

Sum of all diagrams in topology $T_1$ gives
\begin{equation}
  \sum_{i=1}^{16} T_1(i) = \frac{\Gamma(1+3\eps)}{3 \eps} \frac{u_0^6 \ \tau^{-3\eps}}{(4\pi)^{9-3\eps}} \biggl( -\frac{3}{4 \eps^2} - \frac{1.373118(10)}{\eps} + 5.37124(3) \biggr).
\end{equation}

\renewcommand\arraystretch{1.5}
\begin{center}
\begin{longtable}{|c|c|c|c|c|c|}
\caption{Feynman diagrams belonging to the topology $T_{2}$.}  \label{appendixtab:T2} \\

\hline \multicolumn{1}{|c|}{ No.} & \multicolumn{1}{|c|}{ 
%
%
Nickel index \added{with field arguments}
} & \multicolumn{1}{|c|}{ S.F. } & \multicolumn{1}{|c|}{$a_3$} & \multicolumn{1}{|c|}{$a_2$} & \multicolumn{1}{|c|}{$a_1$}
\\ \hline 
\endfirsthead

\multicolumn{4}{|c|}%
{{\bfseries \tablename\ \thetable{} -- continued from previous page}} \\
\hline \multicolumn{1}{|c|}{ No. } &
\multicolumn{1}{|c|}{ 
%
%
Nickel index \added{with field arguments}
} &
\multicolumn{1}{|c|}{ S.F. } & \multicolumn{1}{|c|}{$a_3$} & \multicolumn{1}{|c|}{$a_2$} & \multicolumn{1}{|c|}{$a_1$} \\ \hline 
\endhead

\hline \multicolumn{6}{|c|}{{Continued on next page}} \\ \hline
\endfoot

\hline \hline
\endlastfoot
1 & $e12|23|4|e5|55||:0\psi\_\tpsi\psi\_\tpsi\psi|\tpsi\phi\_\tpsi\phi|\tpsi\psi|0\tpsi\_\psi\tpsi|\tpsi\phi\_\tpsi\psi||$ & $1$ & $-5/768$ & $0.0256619(12)$ & $0.262843(3)$\\ \hline
2 & $e12|23|4|e5|55||:0\psi\_\tpsi\psi\_\tpsi\phi|\tpsi\psi\_\tpsi\phi|\tpsi\psi|0\tpsi\_\psi\tpsi|\tpsi\phi\_\tpsi\psi||$ & $1$ & $-83/2048$ & $0.134695(2)$ & $0.581029(5)$\\ \hline
3 & $e12|23|4|e5|55||:0\psi\_\tpsi\psi\_\tpsi\phi|\tpsi\psi\_\tpsi\psi|\tpsi\psi|0\tpsi\_\phi\tpsi|\tpsi\phi\_\tpsi\psi||$ & $1$ & $63/2048$ & $0.0523358(5)$ & $-0.181356(2)$\\ \hline
4 & $e12|23|4|e5|55||:0\psi\_\tpsi\psi\_\tpsi\psi|\tpsi\phi\_\tpsi\psi|\tpsi\psi|0\tpsi\_\phi\tpsi|\tpsi\phi\_\tpsi\psi||$ & $1$ & $13/1536$ & $0.0167819(1)$ & $-0.0606393(3)$\\ \hline
5 & $e12|23|4|e5|55||:0\psi\_\tpsi\phi\_\tpsi\psi|\psi\tpsi\_\tpsi\psi|\tpsi\psi|0\tpsi\_\phi\tpsi|\tpsi\phi\_\tpsi\psi||$ & $1$ & $39/1024$ & $0.127116(4)$ & $-0.17026(1)$\\ \hline
6 & $e12|23|4|e5|55||:0\psi\_\tpsi\psi\_\tpsi\psi|\phi\tpsi\_\tpsi\phi|\tpsi\psi|0\tpsi\_\psi\tpsi|\tpsi\phi\_\tpsi\psi||$ & $1$ & $-11/768$ & $0.028375(2)$ & $0.108878(12)$\\ \hline
7 & $e12|23|4|e5|55||:0\psi\_\tpsi\psi\_\tpsi\psi|\phi\tpsi\_\tpsi\psi|\tpsi\psi|0\tpsi\_\phi\tpsi|\tpsi\phi\_\tpsi\psi||$ & $1$ & $25/1536$ & $0.040186(1)$ & $-0.013404(11)$\\ \hline
8 & $e12|23|4|e5|55||:0\psi\_\tpsi\phi\_\tpsi\psi|\psi\tpsi\_\tpsi\phi|\tpsi\psi|0\tpsi\_\psi\tpsi|\tpsi\phi\_\tpsi\psi||$ & $1$ & $-49/1024$ & $0.100918(3)$ & $0.407745(14)$\\ \hline

\end{longtable}
\end{center}

Although the topology $T_2$ contains $16$ diagrams in total, results for only $8$ diagrams are stated. That is because the other eight can be shown to give in sum the same contribution as the presented group owing to dual symmetry of the dynamical action (see Sec. \ref{appendixsec:dual} for more details and main article). Hence, contribution from diagrams $T_2(1)$ - $T_2(8)$ should be counted twice.
Sum of all $16$ diagrams in topology $T_2$ thus gives
\begin{equation}
    \sum_{i=1}^{16} T_2(i) = 2  \sum_{i=1}^{8} T_2(i) = \frac{\Gamma(1+3\eps)}{3 \eps} \frac{u_0^6 \ \tau^{-3\eps}}{(4\pi)^{9-3\eps}} \biggl( -\frac{1}{32 \eps^2} + \frac{1.052143(10)}{\eps} + 1.86965(4) \biggr).
\end{equation}
Nickel index representation of diagrams $T_2(9) - T_2(16)$ can be found in Sec. \ref{appendixsec:dual}, Tab. \ref{appendixtab:T2_dual}.

\renewcommand\arraystretch{1.5}
\begin{center}
\begin{longtable}{|c|c|c|c|c|c|}
\caption{Feynman diagrams belonging to the topology $T_{3}$.} \label{appendixtab:T3} \\

\hline \multicolumn{1}{|c|}{ No.} & \multicolumn{1}{|c|}{ 
%
%
Nickel index \added{with field arguments}
} & \multicolumn{1}{|c|}{ S.F. } & \multicolumn{1}{|c|}{$a_3$} & \multicolumn{1}{|c|}{$a_2$} & \multicolumn{1}{|c|}{$a_1$}
\\ \hline 
\endfirsthead

\multicolumn{4}{|c|}%
{{\bfseries \tablename\ \thetable{} -- continued from previous page}} \\
\hline \multicolumn{1}{|c|}{ No. } &
\multicolumn{1}{|c|}{ 
%
%
Nickel index \added{with field arguments}
} &
\multicolumn{1}{|c|}{ S.F. } & \multicolumn{1}{|c|}{$a_3$} & \multicolumn{1}{|c|}{$a_2$} & \multicolumn{1}{|c|}{$a_1$} \\ \hline 
\endhead

\hline \multicolumn{6}{|c|}{{Continued on next page}} \\ \hline
\endfoot

\hline \hline
\endlastfoot
1 & $e12|33|44|5|5|e|:0\tpsi\_\phi\tpsi\_\psi\tpsi|\phi\tpsi\_\psi\tpsi|\phi\tpsi\_\psi\tpsi|\psi\tpsi|\psi\tpsi|0\psi|$ & $1$ & $109/9216$ & $-0.029241(9)$ & $-0.16500(3)$\\ \hline
\end{longtable}
\end{center}

Sum of all diagrams in topology $T_3$ reads
\begin{equation}
   T_3(1) = \frac{\Gamma(1+3\eps)}{3 \eps} \frac{u_0^6 \ \tau^{-3\eps}}{(4\pi)^{9-3\eps}} \biggl( \frac{109}{9216 \eps^2} - \frac{0.029241(9)}{\eps}  -0.16500(3) \biggr).
\end{equation}

\renewcommand\arraystretch{1.5}
\begin{center}
\begin{longtable}{|c|c|c|c|c|c|}
\caption{Feynman diagrams belonging to the topology $T_{4}$.}  \label{appendixtab:T4} \\

\hline \multicolumn{1}{|c|}{ No.} & \multicolumn{1}{|c|}{ 
%
%
Nickel index \added{with field arguments}
} & \multicolumn{1}{|c|}{ S.F. } & \multicolumn{1}{|c|}{$a_3$} & \multicolumn{1}{|c|}{$a_2$} & \multicolumn{1}{|c|}{$a_1$}
\\ \hline 
\endfirsthead

\multicolumn{4}{|c|}%
{{\bfseries \tablename\ \thetable{} -- continued from previous page}} \\
\hline \multicolumn{1}{|c|}{ No. } &
\multicolumn{1}{|c|}{ 
%
%
Nickel index \added{with field arguments}
} &
\multicolumn{1}{|c|}{ S.F. } & \multicolumn{1}{|c|}{$a_3$} & \multicolumn{1}{|c|}{$a_2$} & \multicolumn{1}{|c|}{$a_1$} \\ \hline 
\endhead

\hline \multicolumn{6}{|c|}{{Continued on next page}} \\ \hline
\endfoot

\hline \hline
\endlastfoot
1 & $e12|34|34|5|5|e|:0\tpsi\_\phi\tpsi\_\psi\tpsi|\tpsi\phi\_\psi\tpsi|\psi\tpsi\_\phi\tpsi|\psi\tpsi|\psi\tpsi|0\psi|$ & $1$ & $0$ & $-0.1065288(6)$ & $0.078984(6)$\\ \hline
2 & $e12|34|34|5|5|e|:0\tpsi\_\phi\tpsi\_\psi\tpsi|\phi\tpsi\_\psi\tpsi|\tpsi\psi\_\psi\tpsi|\phi\tpsi|\psi\tpsi|0\psi|$ & $1$ & $0$ & $-0.01277880(15)$ & $0.0191155(4)$\\ \hline
3 & $e12|34|34|5|5|e|:0\tpsi\_\phi\tpsi\_\psi\tpsi|\tpsi\psi\_\psi\tpsi|\psi\tpsi\_\phi\tpsi|\phi\tpsi|\psi\tpsi|0\psi|$ & $1$ & $0$ & $-0.1363848(10)$ & $-1.029619(9)$\\ \hline
4 & $e12|34|34|5|5|e|:0\tpsi\_\phi\tpsi\_\psi\tpsi|\phi\tpsi\_\psi\tpsi|\tpsi\phi\_\psi\tpsi|\psi\tpsi|\psi\tpsi|0\psi|$ & $1$ & $0$ & $-0.00160492(1)$ & $0.00329519(7)$\\ \hline
5 & $e12|34|34|5|5|e|:0\tpsi\_\phi\tpsi\_\psi\tpsi|\tpsi\phi\_\psi\tpsi|\phi\tpsi\_\psi\tpsi|\psi\tpsi|\psi\tpsi|0\psi|$ & $1$ & $0$ & $-0.00160492(1)$ & $-0.0378974(2)$\\ \hline
6 & $e12|34|34|5|5|e|:0\tpsi\_\phi\tpsi\_\psi\tpsi|\phi\tpsi\_\psi\tpsi|\psi\tpsi\_\phi\tpsi|\psi\tpsi|\psi\tpsi|0\psi|$ & $1$ & $0$ & $-0.01562500(11)$ & $-0.0984756(16)$\\ \hline
7 & $e12|34|34|5|5|e|:0\tpsi\_\phi\tpsi\_\psi\tpsi|\phi\tpsi\_\psi\tpsi|\psi\tpsi\_\tpsi\phi|\psi\tpsi|\psi\tpsi|0\psi|$ & $1$ & $0$ & $-0.01277880(13)$ & $0.0175956(5)$\\ \hline
8 & $e12|34|34|5|5|e|:0\tpsi\_\phi\tpsi\_\psi\tpsi|\tpsi\psi\_\psi\tpsi|\phi\tpsi\_\psi\tpsi|\phi\tpsi|\psi\tpsi|0\psi|$ & $1$ & $0$ & $-0.0239526(3)$ & $-0.180845(2)$\\ \hline
9 & $e12|34|34|5|5|e|:0\tpsi\_\phi\tpsi\_\psi\tpsi|\phi\tpsi\_\psi\tpsi|\psi\tpsi\_\tpsi\psi|\psi\tpsi|\phi\tpsi|0\psi|$ & $1$ & $0$ & $-0.0793662(9)$ & $0.038979(3)$\\ \hline
10 & $e12|34|34|5|5|e|:0\tpsi\_\phi\tpsi\_\psi\tpsi|\phi\tpsi\_\psi\tpsi|\phi\tpsi\_\psi\tpsi|\psi\tpsi|\psi\tpsi|0\psi|$ & $1$ & $0$ & $-0.0781250(9)$ & $-0.114523(2)$\\ \hline
\end{longtable}
\end{center}

Sum of all diagrams in topology $T_4$ gives
\begin{equation}
    \sum_{i=1}^{10} T_4(i) = \frac{\Gamma(1+3\eps)}{3 \eps} \frac{u_0^6 \ \tau^{-3\eps}}{(4\pi)^{9-3\eps}} \biggl( \frac{0}{\eps^2} - \frac{0.468750(1)}{\eps}  -1.303391(12) \biggr).
\end{equation}

\begin{center}
\begin{longtable}{|c|c|c|c|c|c|}
\caption{Feynman diagrams belonging to the topology $T_{5}$.} \label{appendixtab:T5} \\

\hline \multicolumn{1}{|c|}{ No.} & \multicolumn{1}{|c|}{ 
%
%
Nickel index \added{with field arguments}
} & \multicolumn{1}{|c|}{ S.F. } & \multicolumn{1}{|c|}{$a_3$} & \multicolumn{1}{|c|}{$a_2$} & \multicolumn{1}{|c|}{$a_1$}
\\ \hline 
\endfirsthead

\multicolumn{4}{|c|}%
{{\bfseries \tablename\ \thetable{} -- continued from previous page}} \\
\hline \multicolumn{1}{|c|}{ No. } &
\multicolumn{1}{|c|}{ 
%
%
Nickel index \added{with field arguments}
} &
\multicolumn{1}{|c|}{ S.F. } & \multicolumn{1}{|c|}{$a_3$} & \multicolumn{1}{|c|}{$a_2$} & \multicolumn{1}{|c|}{$a_1$} \\ \hline 
\endhead

\hline \multicolumn{6}{|c|}{{Continued on next page}} \\ \hline
\endfoot

\hline \hline
\endlastfoot
1 & $e12|34|35|e|55||:0\tpsi\_\phi\tpsi\_\psi\tpsi|\phi\tpsi\_\psi\tpsi|\psi\tpsi\_\tpsi\psi|0\psi|\phi\tpsi\_\psi\tpsi||$ & $1$ & $1/128$ & $0.0336696(9)$ & $-0.040967(3)$\\ \hline
2 & $e12|34|35|e|55||:0\tpsi\_\phi\tpsi\_\psi\tpsi|\psi\tpsi\_\phi\tpsi|\psi\tpsi\_\tpsi\psi|0\psi|\phi\tpsi\_\psi\tpsi||$ & $1$ & $11/2048$ & $-0.0028447(4)$ & $-0.010000(1)$\\ \hline
3 & $e12|34|35|e|55||:0\tpsi\_\phi\tpsi\_\psi\tpsi|\psi\tpsi\_\tpsi\psi|\psi\tpsi\_\phi\tpsi|0\psi|\tpsi\phi\_\tpsi\psi||$ & $1$ & $27/2048$ & $0.030824(1)$ & $-0.050967(9)$\\ \hline
4 & $e12|34|35|e|55||:0\tpsi\_\phi\tpsi\_\psi\tpsi|\psi\tpsi\_\tpsi\psi|\phi\tpsi\_\psi\tpsi|0\psi|\tpsi\phi\_\tpsi\psi||$ & $1$ & $5/1024$ & $-0.081867(6)$ &  $0.57910(10)$ \\ \hline
\end{longtable}
\end{center}

Sum of all diagrams in topology $T_5$ reads
\begin{equation}
    \sum_{i=1}^{4} T_5(i) = \frac{\Gamma(1+3\eps)}{3 \eps} \frac{u_0^6 \ \tau^{-3\eps}}{(4\pi)^{9-3\eps}} \biggl( \frac{1}{32 \eps^2} - \frac{0.020217(7)}{\eps}  +  0.47717(10) \biggr).
\end{equation}

\begin{center}
\begin{longtable}{|c|c|c|c|c|c|}
\caption{Feynman diagrams belonging to the topology $T_{6}$.}
 \label{appendixtab:T6} \\

\hline \multicolumn{1}{|c|}{ No.} & \multicolumn{1}{|c|}{ 
%
%
Nickel index \added{with field arguments}
} & \multicolumn{1}{|c|}{ S.F. } & \multicolumn{1}{|c|}{$a_3$} & \multicolumn{1}{|c|}{$a_2$} & \multicolumn{1}{|c|}{$a_1$}
\\ \hline 
\endfirsthead

\multicolumn{4}{|c|}%
{{\bfseries \tablename\ \thetable{} -- continued from previous page}} \\
\hline \multicolumn{1}{|c|}{ No. } &
\multicolumn{1}{|c|}{ 
%
%
Nickel index \added{with field arguments}
} &
\multicolumn{1}{|c|}{ S.F. } & \multicolumn{1}{|c|}{$a_3$} & \multicolumn{1}{|c|}{$a_2$} & \multicolumn{1}{|c|}{$a_1$} \\ \hline 
\endhead

\hline \multicolumn{6}{|c|}{{Continued on next page}} \\ \hline
\endfoot

\hline \hline
\endlastfoot
1 & $e12|e3|34|5|55||:0\tpsi\_\phi\tpsi\_\psi\tpsi|0\psi\_\tpsi\psi|\psi\tpsi\_\phi\tpsi|\tpsi\psi|\phi\tpsi\_\psi\tpsi||$ & $1$ & $109/18432$ & $-0.020801(3)$ & $-0.128360(8)$\\ \hline
2 & $e12|e3|34|5|55||:0\tpsi\_\psi\tpsi\_\phi\tpsi|0\psi\_\tpsi\psi|\psi\tpsi\_\phi\tpsi|\tpsi\psi|\phi\tpsi\_\psi\tpsi||$ & $1$ & $-17/2304$ & $-0.0101834(1)$ & $-0.0354168(7)$\\ \hline
3 & $e12|e3|34|5|55||:0\tpsi\_\phi\tpsi\_\psi\tpsi|0\psi\_\tpsi\psi|\phi\tpsi\_\psi\tpsi|\tpsi\psi|\phi\tpsi\_\psi\tpsi||$ & $1$ & $-437/18432$ & $0.024175(10)$ & $0.32029(2)$\\ \hline
4 & $e12|e3|34|5|55||:0\tpsi\_\psi\tpsi\_\phi\tpsi|0\psi\_\tpsi\psi|\phi\tpsi\_\psi\tpsi|\tpsi\psi|\phi\tpsi\_\psi\tpsi||$ & $1$ & $79/9216$ & $0.0158157(15)$ & $-0.087571(3)$\\ \hline
\end{longtable}
\end{center}

Sum of all diagrams in topology $T_6$ reads
\begin{equation}
   \sum_{i=1}^{4} T_6(i) = \frac{\Gamma(1+3\eps)}{3 \eps} \frac{u_0^6 \ \tau^{-3\eps}}{(4\pi)^{9-3\eps}} \biggl(  -\frac{17}{1024 \eps^2} + \frac{0.009005(10)}{\eps}  +0.06894(2) \biggr).
\end{equation}

\begin{center}
\begin{longtable}{|c|c|c|c|c|c|}
\caption{Feynman diagrams belonging to the topology $T_{7}$.} \label{appendixtab:T7} \\

\hline \multicolumn{1}{|c|}{ No.} & \multicolumn{1}{|c|}{ 
%
%
Nickel index \added{with field arguments}
} & \multicolumn{1}{|c|}{ S.F. } & \multicolumn{1}{|c|}{$a_3$} & \multicolumn{1}{|c|}{$a_2$} & \multicolumn{1}{|c|}{$a_1$}
\\ \hline 
\endfirsthead

\multicolumn{4}{|c|}%
{{\bfseries \tablename\ \thetable{} -- continued from previous page}} \\
\hline \multicolumn{1}{|c|}{ No. } &
\multicolumn{1}{|c|}{ 
%
%
Nickel index \added{with field arguments}
} &
\multicolumn{1}{|c|}{ S.F. } & \multicolumn{1}{|c|}{$a_3$} & \multicolumn{1}{|c|}{$a_2$} & \multicolumn{1}{|c|}{$a_1$} \\ \hline 
\endhead

\hline \multicolumn{6}{|c|}{{Continued on next page}} \\ \hline
\endfoot

\hline \hline
\endlastfoot
1 & $e12|e3|44|55|5||:0\tpsi\_\psi\tpsi\_\phi\tpsi|0\psi\_\tpsi\psi|\phi\tpsi\_\psi\tpsi|\tpsi\phi\_\tpsi\psi|\psi\tpsi||$ & $1$ & $-209/18432$ & $-0.0591579(1)$ & $0.065316(2)$\\ \hline
2 & $e12|e3|44|55|5||:0\tpsi\_\phi\tpsi\_\psi\tpsi|0\psi\_\tpsi\psi|\phi\tpsi\_\psi\tpsi|\tpsi\phi\_\tpsi\psi|\psi\tpsi||$ & $1$ & $39/2048$ & $0.062287(9)$ & $-0.441410(7)$\\ \hline
\end{longtable}
\end{center}

Sum of all diagrams in topology $T_7$ gives
\begin{equation}
    \sum_{i=1}^{2} T_7(i) = \frac{\Gamma(1+3\eps)}{3 \eps} \frac{u_0^6 \ \tau^{-3\eps}}{(4\pi)^{9-3\eps}} \biggl( \frac{71}{9216 \eps^2} + \frac{0.003129(9)}{\eps} -0.376094(8) \biggr).
\end{equation}

\begin{center}
\begin{longtable}{|c|c|c|c|c|c|}
\caption{Feynman diagrams belonging to the topology $T_{8}$.}  \label{appendixtab:T8} \\

\hline \multicolumn{1}{|c|}{ No.} & \multicolumn{1}{|c|}{ 
%
%
Nickel index \added{with field arguments}
} & \multicolumn{1}{|c|}{ S.F. } & \multicolumn{1}{|c|}{$a_3$} & \multicolumn{1}{|c|}{$a_2$} & \multicolumn{1}{|c|}{$a_1$}
\\ \hline 
\endfirsthead

\multicolumn{4}{|c|}%
{{\bfseries \tablename\ \thetable{} -- continued from previous page}} \\
\hline \multicolumn{1}{|c|}{ No. } &
\multicolumn{1}{|c|}{ 
%
%
Nickel index \added{with field arguments}
} &
\multicolumn{1}{|c|}{ S.F. } & \multicolumn{1}{|c|}{$a_3$} & \multicolumn{1}{|c|}{$a_2$} & \multicolumn{1}{|c|}{$a_1$} \\ \hline 
\endhead

\hline \multicolumn{6}{|c|}{{Continued on next page}} \\ \hline
\endfoot

\hline \hline
\endlastfoot
1 & $e12|e3|45|45|5||:0\tpsi\_\phi\tpsi\_\psi\tpsi|0\psi\_\tpsi\psi|\phi\tpsi\_\psi\tpsi|\tpsi\psi\_\tpsi\psi|\phi\tpsi||$ & $1$ & $-11/1536$ & $0.0176040(16)$ & $0.141967(3)$\\ \hline
2 & $e12|e3|45|45|5||:0\tpsi\_\psi\tpsi\_\phi\tpsi|0\psi\_\tpsi\psi|\phi\tpsi\_\psi\tpsi|\tpsi\psi\_\tpsi\psi|\tpsi\phi||$ & $1$ & $65/3072$ & $0.0460037(3)$ & $-0.027641(2)$\\ \hline
3 & $e12|e3|45|45|5||:0\tpsi\_\phi\tpsi\_\psi\tpsi|0\psi\_\tpsi\psi|\phi\tpsi\_\psi\tpsi|\tpsi\phi\_\tpsi\psi|\psi\tpsi||$ & $1$ & $-11/384$ & $0.056117(5)$ & $0.520326(7)$\\ \hline
4 & $e12|e3|45|45|5||:0\tpsi\_\phi\tpsi\_\psi\tpsi|0\psi\_\tpsi\psi|\phi\tpsi\_\psi\tpsi|\tpsi\psi\_\tpsi\phi|\tpsi\psi||$ & $1$ & $-11/256$ & $-0.009684(7)$ & $1.061565(15)$\\ \hline
5 & $e12|e3|45|45|5||:0\tpsi\_\psi\tpsi\_\phi\tpsi|0\psi\_\tpsi\psi|\phi\tpsi\_\psi\tpsi|\tpsi\phi\_\tpsi\psi|\psi\tpsi||$ & $1$ & $13/768$ & $0.0367653(2)$ & $-0.023507(2)$\\ \hline
6 & $e12|e3|45|45|5||:0\tpsi\_\psi\tpsi\_\phi\tpsi|0\psi\_\tpsi\psi|\phi\tpsi\_\psi\tpsi|\tpsi\psi\_\tpsi\phi|\tpsi\psi||$ & $1$ & $13/512$ & $0.1012803(6)$ & $-0.007610(5)$\\ \hline
7 & $e12|e3|45|45|5||:0\tpsi\_\psi\tpsi\_\phi\tpsi|0\psi\_\tpsi\psi|\phi\tpsi\_\psi\tpsi|\tpsi\psi\_\tpsi\psi|\phi\tpsi||$ & $1$ & $13/3072$ & $0.0092384(1)$ & $-0.0041343(7)$\\ \hline
8 & $e12|e3|45|45|5||:0\tpsi\_\phi\tpsi\_\psi\tpsi|0\psi\_\tpsi\psi|\phi\tpsi\_\psi\tpsi|\tpsi\psi\_\tpsi\psi|\tpsi\phi||$ & $1$ & $-55/1536$ & $0.073721(8)$ & $0.662293(8)$\\ \hline
\end{longtable}
\end{center}

Sum of all diagrams in topology $T_8$ reads
\begin{equation}
    \sum_{i=1}^{8} T_8 (i) = \frac{\Gamma(1+3\eps)}{3 \eps} \frac{u_0^6 \ \tau^{-3\eps}}{(4\pi)^{9-3\eps}} \biggl( -\frac{3}{64 \eps^2} + \frac{0.331047(12)}{\eps}+2.32325(2)\biggr).
\end{equation}

\begin{center}
\begin{longtable}{|c|c|c|c|c|c|}
\caption{Feynman diagrams belonging to the topology $T_{9}$.}  \label{appendixtab:T9} \\

\hline \multicolumn{1}{|c|}{ No.} & \multicolumn{1}{|c|}{ 
%
%
Nickel index \added{with field arguments}
} & \multicolumn{1}{|c|}{ S.F. } & \multicolumn{1}{|c|}{$a_3$} & \multicolumn{1}{|c|}{$a_2$} & \multicolumn{1}{|c|}{$a_1$}
\\ \hline 
\endfirsthead

\multicolumn{4}{|c|}%
{{\bfseries \tablename\ \thetable{} -- continued from previous page}} \\
\hline \multicolumn{1}{|c|}{ No. } &
\multicolumn{1}{|c|}{ 
%
%
Nickel index \added{with field arguments}
} &
\multicolumn{1}{|c|}{ S.F. } & \multicolumn{1}{|c|}{$a_3$} & \multicolumn{1}{|c|}{$a_2$} & \multicolumn{1}{|c|}{$a_1$} \\ \hline 
\endhead

\hline \multicolumn{6}{|c|}{{Continued on next page}} \\ \hline
\endfoot

\hline \hline
\endlastfoot
1 & $e12|34|35|4|5|e|:0\tpsi\_\psi\tpsi\_\phi\tpsi|\psi\tpsi\_\phi\tpsi|\phi\tpsi\_\psi\tpsi|\psi\tpsi|\psi\tpsi|0\psi|$ & $1$ & $-9/1024$ & $0.0073796(3)$ & $-0.035429(3)$\\ \hline
2 & $e12|34|35|4|5|e|:0\tpsi\_\phi\tpsi\_\psi\tpsi|\psi\tpsi\_\phi\tpsi|\psi\tpsi\_\phi\tpsi|\psi\tpsi|\psi\tpsi|0\psi|$ & $1$ & $-27/512$ & $-0.225055(2)$ & $0.720125(15)$\\ \hline
3 & $e12|34|35|4|5|e|:0\tpsi\_\psi\tpsi\_\phi\tpsi|\phi\tpsi\_\psi\tpsi|\psi\tpsi\_\phi\tpsi|\psi\tpsi|\psi\tpsi|0\psi|$ & $1$ & $-3/256$ & $-0.0261171(2)$ & $-0.149277(4)$\\ \hline
4 & $e12|34|35|4|5|e|:0\tpsi\_\phi\tpsi\_\psi\tpsi|\phi\tpsi\_\psi\tpsi|\psi\tpsi\_\phi\tpsi|\psi\tpsi|\psi\tpsi|0\psi|$ & $1$ & $-9/512$ & $-0.0775016(6)$ & $0.308300(6)$\\ \hline
5 & $e12|34|35|4|5|e|:0\tpsi\_\phi\tpsi\_\psi\tpsi|\phi\tpsi\_\psi\tpsi|\phi\tpsi\_\psi\tpsi|\psi\tpsi|\psi\tpsi|0\psi|$ & $1$ & $-15/1024$ & $-0.0121860(6)$ & $0.1197473(15)$\\ \hline
6 & $e12|34|35|4|5|e|:0\tpsi\_\phi\tpsi\_\psi\tpsi|\psi\tpsi\_\phi\tpsi|\phi\tpsi\_\psi\tpsi|\psi\tpsi|\psi\tpsi|0\psi|$ & $1$ & $-45/1024$ & $-0.0115464(14)$ & $0.317614(8)$\\ \hline
7 & $e12|34|35|4|5|e|:0\tpsi\_\psi\tpsi\_\phi\tpsi|\psi\tpsi\_\phi\tpsi|\psi\tpsi\_\phi\tpsi|\psi\tpsi|\psi\tpsi|0\psi|$ & $1$ & $-9/256$ & $-0.0452855(5)$ & $-0.240653(11)$\\ \hline
8 & $e12|34|35|4|5|e|:0\tpsi\_\psi\tpsi\_\phi\tpsi|\phi\tpsi\_\psi\tpsi|\phi\tpsi\_\psi\tpsi|\psi\tpsi|\psi\tpsi|0\psi|$ & $1$ & $-3/1024$ & $0.0001585(2)$ & $-0.0401253(5)$\\ \hline
9 & $e12|34|35|4|5|e|:0\tpsi\_\psi\tpsi\_\phi\tpsi|\psi\tpsi\_\phi\tpsi|\tpsi\psi\_\psi\tpsi|\tpsi\phi|\psi\tpsi|0\psi|$ & $1$ & $0$ & $-0.00301792(1)$ & $-0.0376780(2)$\\ \hline
10 & $e12|34|35|4|5|e|:0\tpsi\_\psi\tpsi\_\phi\tpsi|\phi\tpsi\_\psi\tpsi|\tpsi\psi\_\psi\tpsi|\tpsi\phi|\psi\tpsi|0\psi|$ & $1$ & $-15/256$ & $-0.0207145(14)$ & $0.286976(11)$\\ \hline
11 & $e12|34|35|4|5|e|:0\tpsi\_\phi\tpsi\_\psi\tpsi|\phi\tpsi\_\psi\tpsi|\tpsi\psi\_\psi\tpsi|\tpsi\phi|\psi\tpsi|0\psi|$ & $1$ & $-3/256$ & $0.0105561(4)$ & $0.054546(1)$\\ \hline
12 & $e12|34|35|4|5|e|:0\tpsi\_\phi\tpsi\_\psi\tpsi|\phi\tpsi\_\psi\tpsi|\tpsi\psi\_\psi\tpsi|\tpsi\psi|\phi\tpsi|0\psi|$ & $1$ & $-9/256$ & $-0.0125218(8)$ & $0.281412(2)$\\ \hline
13 & $e12|34|35|4|5|e|:0\tpsi\_\psi\tpsi\_\phi\tpsi|\psi\tpsi\_\phi\tpsi|\tpsi\psi\_\psi\tpsi|\tpsi\psi|\phi\tpsi|0\psi|$ & $1$ & $-9/512$ & $-0.0864748(4)$ & $0.046563(5)$\\ \hline
14 & $e12|34|35|4|5|e|:0\tpsi\_\psi\tpsi\_\phi\tpsi|\phi\tpsi\_\psi\tpsi|\tpsi\psi\_\psi\tpsi|\tpsi\psi|\phi\tpsi|0\psi|$ & $1$ & $-27/512$ & $-0.2668734(12)$ & $0.103043(16)$\\ \hline
15 & $e12|34|35|4|5|e|:0\tpsi\_\phi\tpsi\_\psi\tpsi|\psi\tpsi\_\phi\tpsi|\tpsi\psi\_\psi\tpsi|\tpsi\psi|\phi\tpsi|0\psi|$ & $1$ & $-3/256$ & $-0.0080900(1)$ & $0.1060470(7)$\\ \hline
16 & $e12|34|35|4|5|e|:0\tpsi\_\phi\tpsi\_\psi\tpsi|\psi\tpsi\_\phi\tpsi|\tpsi\psi\_\psi\tpsi|\tpsi\phi|\psi\tpsi|0\psi|$ & $1$ & $0$ & $-0.00301792(1)$ & $0.00728199(6)$\\ \hline
\end{longtable}
\end{center}
In the case of topology $T_9$, we only list $16$ diagrams even though the topology as such contains $32$ Feynman diagrams. The other $16$ diagrams not mentioned need not be calculated owing to dual symmetry (see Sec. \ref{appendixsec:dual} below).

Sum of all $32$ diagrams in this topology thus gives
\begin{equation}
    \sum_{i=1}^{32} T_9 (i) =  2\sum_{i=1}^{16} T_9(i) = \frac{\Gamma(1+3\eps)}{3 \eps} \frac{u_0^6 \ \tau^{-3\eps}}{(4\pi)^{9-3\eps}} \biggl( -\frac{3}{4 \eps^2} - \frac{1.560617(5)}{\eps}+3.69699(4) \biggr).
\end{equation}

Summing contributions from all the topologies we arrive at final result - divergent contributions (in MS scheme) to all 2-point green functions proportional to external frequency

\begin{equation}
   \sum_{\text{3-loop}}= \frac{\Gamma(1+3\eps)}{3 \eps} \frac{u_0^6 \ \tau^{-3\eps}}{(4\pi)^{9-3\eps}} \biggl( -\frac{1581}{1024 \eps^2} - \frac{2.05661(2)}{\eps}+ 11.96278(13)\biggr).
\end{equation}
Nickel index representation of diagrams $T_9(17) - T_9(32)$ can be found in Sec. \ref{appendixsec:dual}, Tab. \ref{appendixtab:T9_dual}.

{\section{Dual symmetry} \label{appendixsec:dual}}

As discussed in the main article, the Feynman diagrams for DyIP process exhibit the dual symmetry
\begin{equation}
   \tpsi(t,\mx) \longleftrightarrow -\varphi(-t,\mx).
   \label{eq:dual_symmetry}
\end{equation}
This reduces the number of diagrams that need to be calculated as it relates certain groups of diagrams - sum of all the diagrams in a group exactly equals sum of diagrams in another. We have identified four such relations (between eight diagram groups) for diagrams in topologies $T_2$ and $T_9$. The fact that these groupings indeed give the same contribution can be proved analytically on the level of integrands before actual integration.

We arrive at the following relations
\begin{align}
\sum_{i=1}^{4} T_2(i) = \sum_{i=9}^{12} T_2(i),\quad \quad
\sum_{i=5}^{8} T_2(i) = \sum_{i=13}^{16} T_2(i),\quad \quad
\sum_{i=1}^8 T_9(i) = \sum_{i=17}^{24} T_9(i),\quad \quad  \sum_{i=9}^{16} T_9(i) = \sum_{i=25}^{32} T_9(i).
\label{appendixeq:dual_reltaions} 
\end{align}
The nickel index representation for diagrams $T_2(9)-T_2(14)$ and $T_9(17)-T_9(32)$ can be found in Tab. \ref{appendixtab:T2_dual} and Tab. \ref{appendixtab:T9_dual} respectively.
\begin{center}
\begin{longtable}{|c|c|}
\caption{Feynman diagrams within topology $T_2$ which need not be calculated as a consequence of dual-symmetry \eqref{eq:dual_symmetry} and relations \eqref{appendixeq:dual_reltaions}.}  \label{appendixtab:T2_dual} \\

\hline \multicolumn{1}{|c|}{ Diagram } & \multicolumn{1}{|c|}{ 
%
%
Nickel index \added{with field arguments}
}
\\ \hline 
\endfirsthead

\multicolumn{2}{|c|}%
{{\bfseries \tablename\ \thetable{} -- continued from previous page}} \\
\hline \multicolumn{1}{|c|}{ Diagram } & \multicolumn{1}{|c|}{ 
%
%
Nickel index \added{with field arguments}
}\\ \hline 
\endhead

\hline \multicolumn{2}{|c|}{{Continued on next page}} \\ \hline
\endfoot

\hline \hline
\endlastfoot

$T_2(9)$ & $e12|23|4|e5|55||:0\tpsi\_\psi\tpsi\_\phi\tpsi|\psi\tpsi\_\phi\tpsi|\psi\tpsi|0\psi\_\tpsi\psi|\phi\tpsi\_\psi\tpsi||$ \\ \hline
$T_2(10)$ & $e12|23|4|e5|55||:0\tpsi\_\phi\tpsi\_\psi\tpsi|\phi\tpsi\_\psi\tpsi|\psi\tpsi|0\psi\_\tpsi\psi|\phi\tpsi\_\psi\tpsi||$ \\ \hline
$T_2(11)$ & $e12|23|4|e5|55||:0\tpsi\_\phi\tpsi\_\psi\tpsi|\psi\tpsi\_\phi\tpsi|\psi\tpsi|0\psi\_\tpsi\psi|\phi\tpsi\_\psi\tpsi||$ \\ \hline
$T_2(12)$ & $e12|23|4|e5|55||:0\tpsi\_\psi\tpsi\_\phi\tpsi|\phi\tpsi\_\psi\tpsi|\psi\tpsi|0\psi\_\tpsi\psi|\phi\tpsi\_\psi\tpsi||$ \\ \hline
$T_2(13)$ & $e12|23|4|e5|55||:0\tpsi\_\psi\tpsi\_\phi\tpsi|\tpsi\phi\_\psi\tpsi|\psi\tpsi|0\psi\_\tpsi\psi|\phi\tpsi\_\psi\tpsi||$ \\ \hline
$T_2(14)$ & $e12|23|4|e5|55||:0\tpsi\_\phi\tpsi\_\psi\tpsi|\tpsi\psi\_\psi\tpsi|\phi\tpsi|0\psi\_\tpsi\psi|\phi\tpsi\_\psi\tpsi||$ \\ \hline
$T_2(15)$ & $e12|23|4|e5|55||:0\tpsi\_\psi\tpsi\_\phi\tpsi|\tpsi\psi\_\psi\tpsi|\phi\tpsi|0\psi\_\tpsi\psi|\phi\tpsi\_\psi\tpsi||$ \\ \hline
$T_2(16)$ & $e12|23|4|e5|55||:0\tpsi\_\phi\tpsi\_\psi\tpsi|\tpsi\phi\_\psi\tpsi|\psi\tpsi|0\psi\_\tpsi\psi|\phi\tpsi\_\psi\tpsi||$ \\ \hline

\end{longtable}
\end{center}
\begin{center}
\begin{longtable}{|c|c|}
\caption{Feynman diagrams within topology $T_9$ which need not be calculated as a consequence of dual-symmetry \eqref{eq:dual_symmetry} and relations \eqref{appendixeq:dual_reltaions}.}  \label{appendixtab:T9_dual} \\

\hline \multicolumn{1}{|c|}{ Diagram } & \multicolumn{1}{|c|}{ 
%
%
Nickel index \added{with field arguments}
}
\\ \hline 
\endfirsthead

\multicolumn{2}{|c|}%
{{\bfseries \tablename\ \thetable{} -- continued from previous page}} \\
\hline \multicolumn{1}{|c|}{ Diagram } & \multicolumn{1}{|c|}{ 
%
%
Nickel index \added{with field arguments}
}\\ \hline 
\endhead

\hline \multicolumn{2}{|c|}{{Continued on next page}} \\ \hline
\endfoot

\hline \hline
\endlastfoot

$T_9(17)$ & $e12|34|35|4|5|e|:0\tpsi\_\psi\tpsi\_\phi\tpsi|\psi\tpsi\_\phi\tpsi|\tpsi\psi\_\psi\tpsi|\phi\tpsi|\psi\tpsi|0\psi|$ \\ \hline
$T_9(18)$ & $e12|34|35|4|5|e|:0\tpsi\_\psi\tpsi\_\phi\tpsi|\phi\tpsi\_\psi\tpsi|\tpsi\psi\_\psi\tpsi|\phi\tpsi|\psi\tpsi|0\psi|$ \\ \hline
$T_9(19)$ & $e12|34|35|4|5|e|:0\tpsi\_\phi\tpsi\_\psi\tpsi|\phi\tpsi\_\psi\tpsi|\tpsi\phi\_\psi\tpsi|\psi\tpsi|\psi\tpsi|0\psi|$ \\ \hline
$T_9(20)$ & $e12|34|35|4|5|e|:0\tpsi\_\psi\tpsi\_\phi\tpsi|\phi\tpsi\_\psi\tpsi|\tpsi\phi\_\psi\tpsi|\psi\tpsi|\psi\tpsi|0\psi|$ \\ \hline
$T_9(21)$ & $e12|34|35|4|5|e|:0\tpsi\_\phi\tpsi\_\psi\tpsi|\psi\tpsi\_\phi\tpsi|\tpsi\phi\_\psi\tpsi|\psi\tpsi|\psi\tpsi|0\psi|$ \\ \hline
$T_9(22)$ & $e12|34|35|4|5|e|:0\tpsi\_\phi\tpsi\_\psi\tpsi|\phi\tpsi\_\psi\tpsi|\tpsi\psi\_\psi\tpsi|\phi\tpsi|\psi\tpsi|0\psi|$ \\ \hline
$T_9(23)$ & $e12|34|35|4|5|e|:0\tpsi\_\psi\tpsi\_\phi\tpsi|\psi\tpsi\_\phi\tpsi|\tpsi\phi\_\psi\tpsi|\psi\tpsi|\psi\tpsi|0\psi|$ \\ \hline
$T_9(24)$ & $e12|34|35|4|5|e|:0\tpsi\_\phi\tpsi\_\psi\tpsi|\psi\tpsi\_\phi\tpsi|\tpsi\psi\_\psi\tpsi|\phi\tpsi|\psi\tpsi|0\psi|$ \\ \hline
$T_9(25)$ & $e12|34|35|4|5|e|:0\tpsi\_\phi\tpsi\_\psi\tpsi|\tpsi\psi\_\psi\tpsi|\psi\tpsi\_\phi\tpsi|\phi\tpsi|\psi\tpsi|0\psi|$ \\ \hline
$T_9(26)$ & $e12|34|35|4|5|e|:0\tpsi\_\psi\tpsi\_\phi\tpsi|\tpsi\psi\_\psi\tpsi|\psi\tpsi\_\phi\tpsi|\phi\tpsi|\psi\tpsi|0\psi|$ \\ \hline
$T_9(27)$ & $e12|34|35|4|5|e|:0\tpsi\_\phi\tpsi\_\psi\tpsi|\tpsi\phi\_\psi\tpsi|\phi\tpsi\_\psi\tpsi|\psi\tpsi|\psi\tpsi|0\psi|$ \\ \hline
$T_9(28)$ & $e12|34|35|4|5|e|:0\tpsi\_\phi\tpsi\_\psi\tpsi|\tpsi\phi\_\psi\tpsi|\psi\tpsi\_\phi\tpsi|\psi\tpsi|\psi\tpsi|0\psi|$ \\ \hline
$T_9(29)$ & $e12|34|35|4|5|e|:0\tpsi\_\psi\tpsi\_\phi\tpsi|\tpsi\psi\_\psi\tpsi|\phi\tpsi\_\psi\tpsi|\phi\tpsi|\psi\tpsi|0\psi|$ \\ \hline
$T_9(30)$ & $e12|34|35|4|5|e|:0\tpsi\_\psi\tpsi\_\phi\tpsi|\tpsi\phi\_\psi\tpsi|\psi\tpsi\_\phi\tpsi|\psi\tpsi|\psi\tpsi|0\psi|$ \\ \hline
$T_9(31)$ & $e12|34|35|4|5|e|:0\tpsi\_\psi\tpsi\_\phi\tpsi|\tpsi\phi\_\psi\tpsi|\phi\tpsi\_\psi\tpsi|\psi\tpsi|\psi\tpsi|0\psi|$ \\ \hline
$T_9(32)$ & $e12|34|35|4|5|e|:0\tpsi\_\phi\tpsi\_\psi\tpsi|\tpsi\psi\_\psi\tpsi|\phi\tpsi\_\psi\tpsi|\phi\tpsi|\psi\tpsi|0\psi|$ \\ \hline

\end{longtable}
\end{center}

All the diagrams in Tabs. \ref{appendixtab:T2_dual}, \ref{appendixtab:T9_dual} have been calculated regardless, using our program implementation. This step was performed as a consistency check of developed software and the results for individual diagram groupings read
\begin{align}
\sum_{i=1}^4 T_2(i) &= \frac{\Gamma(1+3\eps)}{3\eps} \frac{u_0^6 \ \tau_{0}^{-3\eps}}{(4\pi)^{9-3\eps}} \biggl(-\frac{1}{128\eps^{2}} + \frac{0.229474(2) }{\eps} +0.601877(6) \biggr), \label{appendixeq:ress}\\
    \sum_{i=9}^{12} T_2(i) &= \frac{\Gamma(1+3\eps)}{3\eps} \frac{u_0^6 \ \tau_{0}^{-3\eps}}{(4\pi)^{9-3\eps}} \biggl(-\frac{1}{128\eps^{2}} + \frac{0.229475(5) }{\eps} +0.60188(2) \biggr).\\
\sum_{i=5}^8 T_2(i) &= \frac{\Gamma(1+3\eps)}{3\eps} \frac{u_0^6 \ \tau_{0}^{-3\eps}}{(4\pi)^{9-3\eps}} \biggl(-\frac{1}{128\eps^{2}}  + \frac{0.296596(6) }{\eps}  +0.33295(2)\biggr) \\
    \sum_{i=13}^{16} T_2 (i) &= \frac{\Gamma(1+3\eps)}{3\eps} \frac{u_0^6 \ \tau_{0}^{-3\eps}}{(4\pi)^{9-3\eps}} \biggl(-\frac{1}{128\eps^{2}}  + \frac{0.296597(7)}{\eps} + 0.33295(4)\biggr),  \\
\sum_{i=1}^8 T_9(i) &= \frac{\Gamma(1+3\eps)}{3\eps} \frac{u_0^6 \ \tau_{0}^{-3\eps}}{(4\pi)^{9-3\eps}} \biggl(-\frac{3}{16 \eps^{2}} - \frac{0.390154(2)}{\eps}+ 1.00030(2) \biggr),\\
    \sum_{i=17}^{24} T_9 (i) &= \frac{\Gamma(1+3\eps)}{3 \eps} \frac{u_0^6 \ \tau_{0}^{-3\eps}}{(4\pi)^{9-3\eps}} \biggl(-\frac{3}{16\eps^{2}}  - \frac{0.390154(4)}{\eps} + 1.00030(2) \biggr), \\
\sum_{i=9}^{16} T_9(i) &= \frac{\Gamma(1+3\eps)}{3\eps} \frac{u_0^6 \ \tau_{0}^{-3\eps}}{(4\pi)^{9-3\eps}} \biggl(-\frac{3}{16 \eps^{2}} - \frac{0.390154(2)}{\eps}  + 0.84819(2)\biggr),\\
    \sum_{i=25}^{32} T_9 (i) &= \frac{\Gamma(1+3\eps)}{3\eps} \frac{u_0^6 \ \tau_{0}^{-3\eps}}{(4\pi)^{9-3\eps}} \biggl(-\frac{3}{16\eps^{2}}  - \frac{0.390154(4) }{\eps} +0.848192(17) \biggr),\\
\label{appendixeq:resf}
\end{align}
It is clear from the results 
\eqref{appendixeq:ress} - \eqref{appendixeq:resf} that the respective groups indeed sum to the same values up to error of numerical method. 

{\section{Results for renormalization constants $Z_2$, $Z_3$} \label{appendix:other_constants}}
As mentioned in the main article, we did run computations also on $2$-pt. diagrams contributing to renormalization constants $Z_2$, $Z_3$. 
For these analytical results up to three-loop order exist \cite{bonfim1980} as these can be obtained from simpler quazi-static limit (Potts model in the single-state limit).
We can thus proceed to compute these constants numerically from the full dynamical action using the developed programs as a means of consistency check. The RG constants $Z_2$ and $Z_3$ are calculated parts of 1PI Green function $\Gamma^{(1,1)}$ proportional to external momentum ${\bm p}^2$ and parameter $\tau_0$, respectively. Similarly to the calculation of $Z_1$, we have to analyze the same $93$ three-loop diagrams in $9$ distinct topologies as presented in Sec. \ref{appendix:diagram}. However, different symbolical manipulations must be performed within our program. Since this is not the main result of the paper we will not list all intermediate results here and only state the final expressions. In MS scheme we have obtained RG constants $Z_2$ and $Z_3$ 
\begin{align}
	Z_2 &= 1 + \frac{g}{2\eps} \frac{1}{6} +  \frac{g^2}{(2\eps)^2} \frac{11}{36} + \frac{g^2}{2\eps} \biggl( \frac{-37}{432} \biggr)
    + \frac{g^3}{(2\eps)^3}\biggl(\frac{473}{648}\biggr) + \frac{g^3}{(2\eps)^2} \biggl(-0.7318(1)\biggr) + \frac{g^3}{(2\eps)} \biggl(0.1469(2)\biggr),  \label{appendixeq:Z2}\\
    Z_3 &= 1+\frac{g}{2 \eps} + \frac{g^2}{(2\eps)^2} \frac{9}{4} + \frac{g^2}{2\eps} \biggl( \frac{-47}{48} \biggr) + \frac{g^3}{(2\eps)^3}\biggl(6\biggr) + \frac{g^3}{(2\eps)^2} \biggl(-6.3703(2)\biggr) + \frac{g^3}{(2\eps)} \biggl(4.060(6)\biggr), \label{appendixeq:Z3}
\end{align}
To avoid unnecessary long calculations, these constants $Z_2$ and $Z_3$ were calculated with a smaller precision than $Z_1$.
Nonetheless, the obtained expressions using our semi-analytical approach for constants $Z_2$, $Z_3$ are in accordance with analytic results obtained in \cite{bonfim1980} which read
\begin{align}
	Z_{2,\ \cite{bonfim1980}} &= 1 + \frac{g}{2\eps} \frac{1}{6} +  \frac{g^2}{(2\eps)^2} \frac{11}{36} + \frac{g^2}{2\eps} \biggl( \frac{-37}{432} \biggr)
    + \frac{g^3}{(2\eps)^3}\biggl(\frac{473}{648}\biggr) + \frac{g^3}{(2\eps)^2} \biggl(-0.7318\biggr) + \frac{g^3}{(2\eps)} \biggl(0.1470\biggr),  \label{appendixeq:Z2_bon}\\
    Z_{3,\ \cite{bonfim1980}} &= 1+\frac{g}{2 \eps} + \frac{g^2}{(2\eps)^2} \frac{9}{4} + \frac{g^2}{2\eps} \biggl( \frac{-47}{48} \biggr) + \frac{g^3}{(2\eps)^3}\biggl(6\biggr) + \frac{g^3}{(2\eps)^2} \biggl(-6.3703\biggr) + \frac{g^3}{(2\eps)} \biggl(4.063\biggr). \label{appendixeq:Z3_bon}
\end{align}
Note, that these analytic expressions were written in decimal format and truncated for ease of comparison.

In the case of RG constant $Z_4$ the situation becomes more involved. The main reason is that analysis of the Green function $\Gamma^{(2,1)}$ requires a calculation of total $957$ such diagrams in $17$ topologies within the three-loop order. The calculation of $Z_4$ from full dynamical action thus represents serious computational load and was not performed. 

\bibliographystyle{apsrev}
\bibliography{mybib_supp}